\documentclass[conference]{IEEEtran}

\usepackage{graphicx}
\usepackage{amsmath}
\usepackage{listings}
\usepackage[table]{xcolor}
\usepackage{multirow,colortbl}
\usepackage{hyperref}
\usepackage{cleveref}
\usepackage[USenglish]{babel}
\usepackage{booktabs}
\usepackage{tikz,pgf-pie,wheelchart}
\usepackage{xurl}
\usepackage{stfloats}
\usepackage[acronym]{glossaries}
\newacronym{cpu}{CPU}{Central Processing Unit}
\newacronym{gpu}{GPU}{Graphics Processing Unit}
\newacronym{fpga}{FPGA}{Field-Programmable Gate Array}
\newacronym{asic}{ASIC}{Application-Specific Integrated Circuit}
\newacronym{os}{OS}{Operating System}
\newacronym{api}{API}{Application Programming Interface}
\newacronym{ssh}{SSH}{Secure Shell}
\newacronym{uart}{UART}{Universal Asynchronous Receiver-Transmitter}
\newacronym{lan}{LAN}{Local Area Network}
\newacronym{nat}{NAT}{Network Address Translation}
\newacronym{vlan}{VLAN}{Virtual Local Area Network}
\newacronym{isp}{ISP}{Internet Service Provider}
\newacronym{tcp}{TCP}{Transmission Control Protocol}
\newacronym{http}{HTTP}{Hypertext Transfer Protocol}
\newacronym{https}{HTTPS}{Hypertext Transfer Protocol Secure}
\newacronym{ssl}{SSL}{Secure Sockets Layer}
\newacronym{tls}{TLS}{Transport Layer Security}
\newacronym{mitm}{MITM}{Man-In-The-Middle}
\newacronym{rce}{RCE}{Remote Code Execution}
\newacronym{cve}{CVE}{Common Vulnerabilities and Exposures}
\newacronym{sd}{SD}{Secure Digital}
\newacronym{emmc}{eMMC}{Embedded Multimedia Card}
\newacronym{fsbl}{FSBL}{First Stage Boot Loader}

\glsdisablehyper
\usepackage{tabularx}
\usepackage{url}

\definecolor{archBootText}{RGB}{120,90,150}
\definecolor{archLibText}{RGB}{40,110,140}
\definecolor{archMinerText}{RGB}{60,120,60}
\definecolor{archWebText}{RGB}{40,80,160}
\definecolor{archFWText}{RGB}{150,60,60}
\definecolor{archPhysText}{RGB}{120,80,120}
\definecolor{archDebugText}{RGB}{180,120,60}
\definecolor{archConsoleText}{RGB}{90,90,90}

\definecolor{LightGreen}{RGB}{170,255,140}
\definecolor{LightRed}{RGB}{255,80,80}

\definecolor{pastelRed}{RGB}{220,90,90}
\definecolor{pastelOrange}{RGB}{230,150,80}
\definecolor{pastelYellow}{RGB}{220,190,60}
\definecolor{pastelBlue}{RGB}{120,160,220}

\definecolor{bluePierre}{HTML}{004586}
\definecolor{redPierre}{HTML}{ff420e}
\definecolor{yellowPierre}{HTML}{ffd320}
\definecolor{greenPierre}{HTML}{579d1c}
\definecolor{brownPierre}{HTML}{7e0021}

\title{Firmware Distribution as Attack Surface:\\
A Security Study of ASIC Cryptocurrency Miners}

\author{
\IEEEauthorblockN{
Pierre Pouliquen\IEEEauthorrefmark{1}\IEEEauthorrefmark{2},
Hadrien Barral\IEEEauthorrefmark{3},
David Naccache\IEEEauthorrefmark{1},
Thibaut Heckmann\IEEEauthorrefmark{2},
Antoine Houssais\IEEEauthorrefmark{2}
}
\IEEEauthorblockA{\IEEEauthorrefmark{1}
DI ENS, École normale supérieure, PSL, CNRS, 75005 Paris, France\\
\texttt{firstname.lastname@ens.psl.eu}}
\IEEEauthorblockA{\IEEEauthorrefmark{2}
Laboratoire de Recherche et d'Innovation, Gendarmerie Nationale, Chaire HUNUM, 77000 Melun, France\\
\texttt{firstname.lastname@gendarmerie.interieur.gouv.fr}}
\IEEEauthorblockA{\IEEEauthorrefmark{3}
Univ Gustave Eiffel, CNRS, LIGM, 77454 Marne-la-Vallée, France\\
\texttt{firstname.lastname@univ-eiffel.fr}}
}
\begin{document}

\maketitle

\pagestyle{plain}

\begin{abstract}
ASIC cryptocurrency miners are a core component of blockchain infrastructures, directly converting computation and energy into monetary value. Despite their economic importance, their security is rarely evaluated in a structured manner. In this paper, we show that the firmware distribution ecosystem of mining devices fundamentally challenges existing trust assumptions. We introduce a scalable methodology based on the collection and static analysis of publicly distributed firmware artifacts, requiring neither device access nor runtime interaction. Applying this approach, we reconstruct and analyze 134 firmware images spanning manufacturers that account for over 99\% of deployed miners (Bitmain, MicroBT, Canaan, Iceriver). Our results reveal that firmware artifacts alone are sufficient to recover internal architecture, identify security weaknesses, and reconstruct complete attack paths leading to high-impact adversarial objectives. In particular, our analysis reveals vulnerabilities that enable realistic large-scale attack scenarios, including firmware phishing and the exploitation of miners still operating over Stratum V1. Validation on two real devices confirms that publicly distributed artifacts closely reflect deployed software and that these weaknesses translate into attack capabilities. Overall, our study shows that firmware distribution mechanisms themselves constitute a primary attack surface, significantly lowering the barrier to compromise in the ASIC mining ecosystem.
\end{abstract}

\section{Introduction}
Cryptocurrencies are decentralized digital monetary systems in which transactions are recorded on a public blockchain and validated collectively by network participants rather than by a trusted central authority. Since the introduction of Bitcoin, these systems have evolved from a niche technological proposal into a large-scale financial and computational ecosystem. This sustained relevance is reflected both in the scientific literature and in economic indicators. A recent bibliometric analysis identifies more than 41\,000 scientific publications related to blockchain technologies between 2008 and 2023, with over 60\% of them published during the 2021--2023 period, indicating a persistent and growing research interest in cryptocurrencies and their underlying infrastructures \cite{bao2025bibliometricanalysisscientificpublications,bonneau2015sok}. In parallel, the total market capitalization of all cryptocurrencies reached approximately \$3\,trillion USD in March/April~2024, with the largest asset (Bitcoin) alone accounting for roughly \$1.7--\$1.8\,trillion USD at that time, underscoring the continued economic significance of this asset class~\cite{ZHANG2024106114}.
Modern cryptocurrency miners are no longer general purpose computers but networked embedded systems operated continuously in production environments and managed remotely at scale. Their firmware governs parameters that directly influence revenue, such as mining pool configuration, payout addresses, and operating frequency, as well as those related to safe operation, including power delivery and thermal management. Firmware correctness and integrity therefore directly affect both economic outcomes and operational stability.
In parallel, the \gls{asic} mining hardware market is dominated by a small number of manufacturers \cite{CambridgeMining2025}, whose design choices and firmware update mechanisms are deployed across a large fraction of the global hashrate. This strong vendor concentration implies that weaknesses in miner firmware do not remain confined to individual devices but can propagate in the ecosystem, affecting entire blockchain networks. Consequently, the security of cryptocurrencies now depends not only on cryptographic protocols, but also on the robustness of the firmware running on a limited set of mining platforms \cite{BabkinFirmWar2023}. 

\subsection{Our contribution}

We analyze the \gls{asic} mining ecosystem by combining a market study of the dominant manufacturers with a security analysis of their products. Our approach relies on a reproducible methodology based on the systematic collection and static analysis of publicly distributed update firmware artifacts. Applied at scale, this method enables the recovery of detailed architectural and security relevant information without physical access to devices or dynamic interaction. Using this approach, we identify major vulnerabilities affecting real-world miners and show that firmware update packages alone expose a substantial attack surface and a large amount of sensitive operational information, significantly lowering the cost of attacks since it is not necessary to own the miner to know its vulnerabilities.

\subsection{Paper organization}

We begin by introducing the cryptocurrency mining ecosystem and its security implications (\Cref{sec:background}). We then describe the firmware architecture and its main attack surfaces, with a focus on the update workflow as a central trust boundary (\Cref{sec:arch}, \Cref{sec:firmupd}).

Next, we formalize an attacker model (\Cref{sec:attack}) and present its operational realization as an attack pipeline from access to impact (\Cref{sec:attacker_model}).

We then introduce our large-scale firmware analysis methodology (\Cref{sec:methodo}), show the information that can be extracted (\Cref{sec:inforetri}), and derive realistic attack scenarios (\Cref{sec:security_exp}).

Finally, we validate our findings (\Cref{sec:validation}), discuss mitigations (\Cref{sec:sugg}), and conclude (\Cref{sec:conclusion}).

\subsection{Related work}
\label{sec:related}

Prior work follows two largely independent directions: large-scale firmware analysis and cryptocurrency mining security.

Firmware analysis has shown that publicly available images enable scalable vulnerability discovery without device access~\cite{CostinUsenix14}, later extended with dynamic and emulations techniques such as Firmadyne, Firm-AFL, and P2IM~\cite{costin2016dynamic,BaFIRMAFLUsenix19,FengP2IMUsenix20}. Surveys further systematize extraction methods and vulnerability classes in embedded firmware~\cite{ulhaq2023firmware,bakhshi2024iotfirmwarevulnerabilities}.

In parallel, mining security research focuses on protocol and network-level attacks, including Stratum weaknesses, hashrate redirection, and routing attacks~\cite{recabarren2017hardeningstratum,liu2021disappeared,StratumMITM,StratumAttackSurface,Tran2024RoutingAttacksMiningPools}.

However, these directions remain disconnected: firmware studies are domain-agnostic, while mining security works do not consider firmware and update mechanisms. As summarized in \Cref{app:related_extended}, no prior work combines cross-vendor firmware analysis with mining-specific attack modeling. Our work bridges this gap by linking firmware artifacts to concrete attacker capabilities and economic impact.

\section{What is cryptocurrency and why do we need mining machines?}\label{sec:background}

Cryptocurrencies originate from early work on cryptographically enforced electronic cash. In 1983, in his seminal paper \emph{Blind Signatures for Untraceable Payments}, Chaum introduced the first formal construction of a digital payment system based on cryptographic primitives, enabling privacy-preserving electronic transactions mediated by a trusted authority \cite{Chaum1983BlindSignatures}. While these early proposals relied on centralized intermediaries, they established the foundational concepts of cryptographic money.

Building on these ideas, Bitcoin, proposed by Nakamoto in 2008 \cite{nakamoto2008bitcoin}, was the first system to maintain a consistent and antifraud transaction ledger in an open network without any trusted central authority. It combined public-key cryptography, hash functions, and a decentralized consensus mechanism based on proof-of-work, enabling mutually distrustful participants to agree on transaction validity.

Subsequent systems such as Ethereum \cite{Buterin2014Ethereum} generalized this model by introducing programmable smart contracts while initially retaining proof-of-work as a consensus mechanism. This further reinforced the economic role of mining infrastructures beyond simple transaction validation.

In these systems, ownership of funds is defined by cryptographic keys, and transactions are broadcast to a peer-to-peer network before being incorporated into blocks through collective validation. Miners emerged as a direct consequence of this design, assembling transactions into blocks and competing to solve proof-of-work challenges that are computationally costly to generate but easy to verify, thereby securing the ledger against adversarial manipulation.

In practice, most miners do not operate independently but participate in \emph{mining pools}, which coordinate the work of large numbers of devices. A pool assigns hashing jobs to connected miners, collects submitted shares as proof of contributed work, and distributes rewards proportionally. As a result, miners continuously maintain network connections to a small set of pool endpoints, making pool configuration parameters (such as pool addresses and payout identifiers) a central economic control surface.

Early mining was performed on general-purpose \gls{cpu}s, then on \gls{gpu}s, and later on \gls{fpga}s. As competition intensified, mining evolved into a hardware arms race, leading to the development of \acrfull{asic}s optimized exclusively for hash computation~\cite{wang2021asic}. These machines exist solely to perform a cryptographic function at maximum efficiency. Modern cryptocurrency mining is therefore inseparable from specialized hardware and industrial deployment \cite{CambridgeMining2025}.

This evolution has produced a mining ecosystem that is decentralized at the protocol level but highly centralized at the hardware level. The global \gls{asic} market is overwhelmingly dominated by a small number of manufacturers, primarily Bitmain (China) \cite{BitmainSite}, MicroBT (China) \cite{MicroBTSite}, and Canaan (China) \cite{CanaanSite}, which together supply more than 99\% of deployed mining devices \cite{CambridgeMining2025}. As a result, the security properties of cryptocurrencies now depend not only on cryptographic protocols, but also on the correctness, integrity, and update mechanisms of the firmware running on a limited set of mining platforms. Vulnerabilities at this layer can affect a significant fraction of the global hashrate and therefore constitute a systemic risk for the underlying blockchain networks.

\section{Software architecture of a cryptocurrency miner firmware}\label{sec:arch}

\begin{figure}[t]
  \centering
  \includegraphics[width=\columnwidth,trim={0 0.90cm 0 1.0cm},clip]{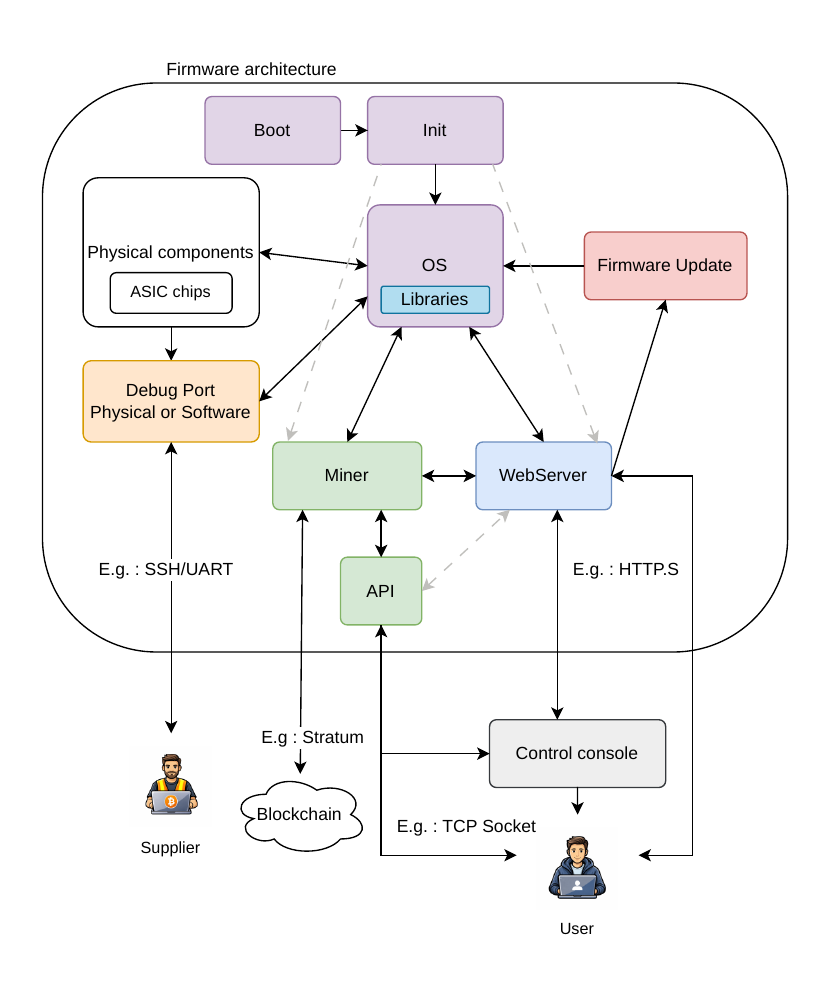}
  \caption{Layered software architecture of a cryptocurrency mining firmware.}
  \label{fig:firmware_arch}
\end{figure}

\Cref{fig:firmware_arch} presents a functional architectural abstraction of a cryptocurrency miner firmware, designed to capture the organization of the software stack rather than provide an exhaustive description. After \emph{\textcolor{archBootText}{Boot}} and hardware initialization via \emph{\textcolor{archBootText}{Init}}, a lightweight embedded \emph{\textcolor{archBootText}{Operating System}} (OS) and its \emph{\textcolor{archLibText}{Libraries}}, most commonly Linux, are loaded and provide core services such as process management, networking, storage, and access to hardware drivers.

The mining application (\emph{\textcolor{archMinerText}{Miner}}) runs on top of this \emph{\textcolor{archBootText}{OS}} and implements the proof-of-work logic. It communicates with mining pools (\emph{Blockchain}) using Stratum, the standard protocol through which miners receive work (hashing jobs) and submit computed shares. It then distributes work to the \gls{asic} chips (\emph{\textcolor{archPhysText}{Physical components}}) through \emph{\textcolor{archBootText}{OS}}  control paths, collects results, and controls parameters including frequency, voltage, and thermal limits.

In parallel, management components implement a management plane that either directly exposes a configuration \emph{\textcolor{archMinerText}{API}} within the mining software, or relies on an HTML administrative interface provided by a \emph{\textcolor{archWebText}{WebServer}}, which forwards configuration commands to the mining process through internal control channels. These management interfaces are accessible to the operator either programmatically or through a browser, optionally via a \emph{\textcolor{archConsoleText}{Control console}}. The \emph{\textcolor{archFWText}{Firmware Update}} mechanism interacts with the operating system, the miner, and the web interface to define how software components can be replaced or modified over time, thereby shaping the update trust model and long-term security posture of the device. In addition, miners may expose a \emph{\textcolor{archDebugText}{Debug port}}, either logically through remote access mechanisms such as \acrshort{ssh} or physically through interfaces like \gls{uart}, primarily intended for maintenance or vendor-side diagnostics.

This architectural abstraction is sufficient to reason about miner behavior, control flows, update paths, and security-relevant attack surfaces. The different elements are detailed in \Cref{sec:minerarch} in the appendix.

\section{Firmware update workflow as a primary security entry point}
\label{sec:firmupd}

Cryptocurrency mining devices rely on a manual firmware update workflow under the direct control of the operator. Unlike consumer IoT ecosystems, miners do not implement automatic update polling or enforced over-the-air patching. Updates are explicitly initiated through privileged management interfaces. Architecturally, this workflow corresponds to the \emph{\textcolor{archFWText}{Firmware Update}} component shown in \Cref{fig:firmware_arch}, which forms the trust boundary between the external management plane and internal system components (\emph{\textcolor{archBootText}{OS}}, \emph{\textcolor{archMinerText}{Miner}}, and \emph{\textcolor{archLibText}{Libraries}}). Importantly, this same update path is also the primary observation point of our methodology.

In practice, users download update packages from public vendor repositories, connect to the miner administration interface (embedded web interface or vendor management software), and upload the package for installation. Although these packages are often partial and do not constitute a complete system image, they remain sufficiently representative to expose the internal software structure, update logic, and trust boundaries of the device.

A key property of this ecosystem is that these firmware packages are publicly accessible and can be collected at scale without requiring access to physical hardware. This enables offline analysis capable of recovering system architecture, identifying exposed services, and understanding security-critical mechanisms, as these artifacts reflect the structural elements governing the deployed system.

As a result, static analysis of these artifacts enables extensive offline reverse engineering, revealing operating system components, boot and update mechanisms, management services, authentication logic, and, in many cases, exploitable vulnerabilities. This asymmetry is exacerbated by the fact that firmware analysis is largely automated and low-cost, enabling attackers to prepare attacks at scale without facing network or operational constraints and without acquiring the hardware.

This openness is not incidental but largely driven by economic constraints. Mining hardware profitability degrades rapidly over time, typically over weeks to months due to increasing network difficulty and hardware turnover, forcing operators to maximize short-term returns through overclocking and performance tuning \cite{10.1145/3581971.3581978}. This requirement encourages the use of modified or unofficial firmware. Vendors therefore expose or tolerate user control over performance parameters, which amplifies the security impact of weaknesses in update verification mechanisms.

\section{Attack objectives and adversarial model}
\label{sec:attack}
Cryptocurrency mining devices constitute high-value targets for adversaries because they operate continuously, expose network services by design, and directly convert computation and energy into monetary value \cite{enisa2019_threat_landscape_2018}. Unlike most consumer IoT devices, whose compromise yields only indirect benefits, a miner produces a steady and measurable revenue stream as long as it remains operational. This economic asymmetry helps explain why illicit mining and cryptojacking have become persistent threat categories in the wild \cite{kaspersky2022_crypto_miners_on_the_rise,merces2018trendmicro_iot_miner,zimba2021demystifyingcryptocurrencyminingattacks,sari2017exploitingcryptocurrencyminers,bakhshi2024iotfirmwarevulnerabilities}.

In this section, we focus on the objectives that an adversary may pursue once a mining device is compromised. Rather than describing how attacks are carried out, we characterize the impact of a compromise in terms of attacker goals and observable outcomes. This perspective allows us to reason about security at the level of consequences, independently of specific vulnerabilities or exploitation paths.

\textbf{Full device takeover and infrastructure abuse.}
An attacker may seek full control of the miner in order to repurpose it as a general-purpose networked system. Once compromised, miners can be integrated into botnets, used for DDoS attacks, or abused as proxy and relay nodes. This type of outcome is consistent with the broader history of embedded-device compromise, as illustrated by large-scale botnets such as \emph{Mirai} and more recent Linux malware campaigns targeting network-connected devices \cite{mirai_cloudflare,condibot_scmedia}. In the mining ecosystem itself, prior reports and public disclosures have documented remote compromise, malicious code deployment, and persistent abuse of ASIC devices, making such scenarios realistic in practice \cite{queenant2016,antbleed2017,dcentral2025,chambers2022}.

\textbf{Cryptocurrency theft and revenue redirection.}
A central objective consists in diverting mining revenue without disrupting apparent device availability. This can be achieved by modifying wallet addresses, mining pool endpoints, or Stratum parameters, either locally through configuration tampering or firmware modification, or remotely via man-in-the-middle attacks on unauthenticated mining protocols. Prior work has shown that mining with Stratum V1 remains vulnerable to stealthy hashrate theft, share manipulation, and proxy redirection attacks \cite{liu2021disappeared,recabarren2017hardeningstratum,StratumMITM,StratumAttackSurface}. At the infrastructure level, routing manipulation can transparently redirect miners toward attacker pools without requiring local compromise \cite{Tran2024RoutingAttacksMiningPools,bgp_hijack_mining}. Similar behavior has also been observed in malware targeting mining software, where wallet addresses are replaced to redirect rewards \cite{satori_coin_robber}. These attacks are particularly effective because they preserve normal device operation while silently redirecting profits.

\textbf{Physical degradation and hardware damage.}
Beyond direct revenue theft, attackers may target the physical integrity and lifetime of mining hardware. By manipulating voltage, frequency, fan control, or thermal management parameters, an adversary can induce overheating, chronic instability, or accelerated component aging. This scenario is particularly relevant in mining environments because performance tuning and clock-rate adjustment are already central operational controls \cite{10.1145/3581971.3581978}. In addition, public reports on malware targeting ASIC devices indicate that hostile access to these parameters is not purely theoretical and may threaten device stability or safe operating conditions \cite{hant_malware,dcentral2025}.

\textbf{Performance degradation and operational disruption.}
Finally, adversaries may seek to degrade mining performance without achieving persistent control. Network flooding, repeated service restarts, denial-of-service attacks, or resource exhaustion can significantly reduce hashrate and availability while remaining difficult to attribute. Such attacks have been observed in practice, notably through DDoS campaigns targeting mining pools and related infrastructure \cite{dd4bc_coindesk}. More generally, mining activity remains exposed to routing-level and network-level disruptions that can affect effective work delivery and reward collection even without full compromise of the miner itself \cite{Tran2024RoutingAttacksMiningPools,enisa2019_threat_landscape_2018}.

\begin{figure*}[t]
  \centering
  \includegraphics[width=1\textwidth]{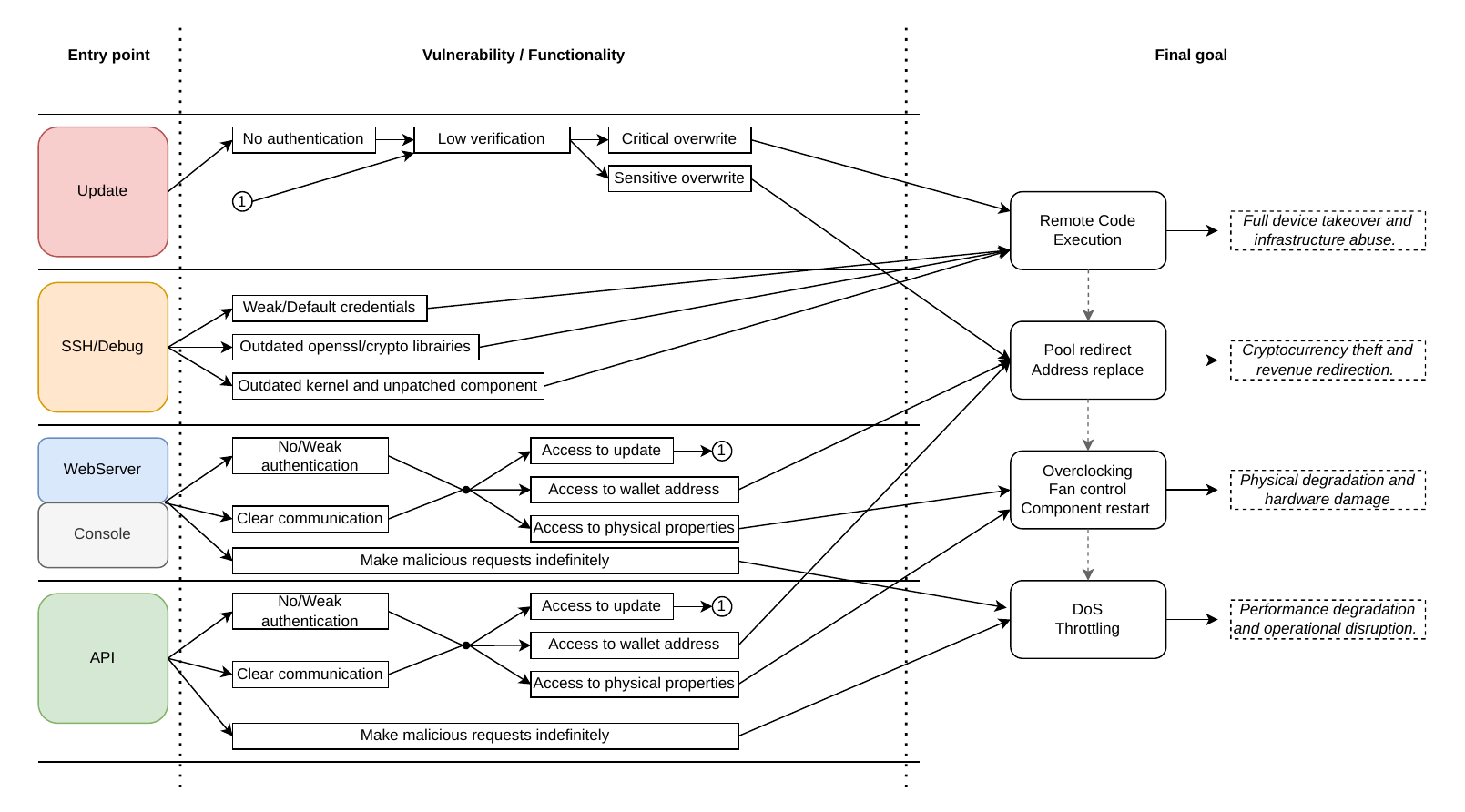}
   \caption{Structured view of the attack pipeline linking entry points, vulnerabilities, capabilities, and attacker objectives. This figure does not aim to exhaustively enumerate all possible paths, but rather to illustrate representative vulnerability chains in order to provide an intuitive understanding of how different attack scenarios can lead to concrete objectives.}
  \label{fig:attack_model}
\end{figure*}

\section{Attacker model and attack pipeline}
\label{sec:attacker_model}

This section provides an operational view of how the objectives described in \Cref{sec:attack} can be achieved in practice. We describe how an adversary progresses from initial access to concrete impact by interacting with the mining device through its exposed attack surface.

Mining devices are typically deployed within private networks and are not directly exposed to the Internet. As a result, an attacker must first gain a foothold in the local environment in which miners operate. This initial access constitutes a necessary precondition for most realistic attacks.

\begin{figure}[ht!]
  \centering
  \includegraphics[width=\columnwidth]{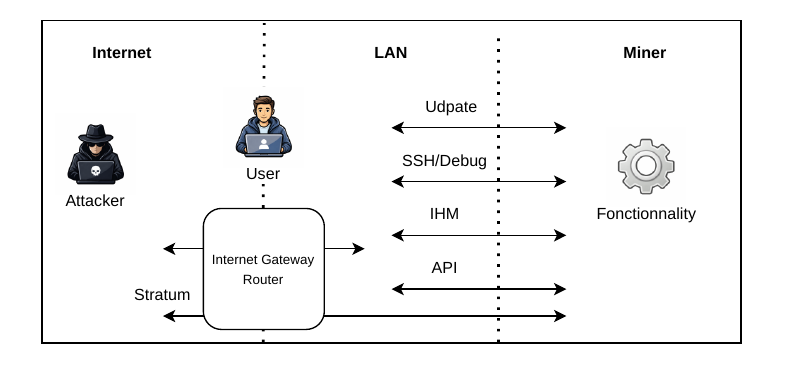}
  \caption{Attacker perspective: gaining access to the miner network and positioning within the local environment through different compromise vectors.}
  \label{fig:attacker_perspective}
\end{figure}

As illustrated in \Cref{fig:attacker_perspective}, this can be achieved through several practical vectors, including network-level attacks (e.g., interception or manipulation of mining traffic), social engineering targeting operators (e.g., malicious firmware distribution), or compromise of adjacent infrastructure such as routers or management hosts.

Once this foothold is obtained, the attacker can interact with the device through a limited set of entry points defined by its architecture (\Cref{sec:arch}), namely firmware update mechanisms, debug interfaces such as SSH, embedded management interfaces, and exposed APIs. These entry points define the effective attack surface available in practice.

The attack can then be modeled as a progressive pipeline:

\begin{center}
Access $\rightarrow$ Entry point $\rightarrow$ Vulnerability $\rightarrow$ Capability $\rightarrow$ Objective
\end{center}

\Cref{fig:attack_model} summarizes the global relationships between entry points, vulnerabilities, capabilities, and final objectives. It highlights how each stage of the attack constrains the next, and how different attack paths may converge toward similar capabilities and outcomes. More precisely, entry points determine which vulnerabilities are reachable, vulnerabilities determine which functionalities can be accessed, and these functionalities translate into attacker capabilities such as firmware modification, configuration control, or access to physical parameters. These capabilities ultimately enable the objectives described in \Cref{sec:attack}. For completeness, the attacker workflow can also be represented as a temporal pipeline emphasizing execution steps rather than structural relationships. This alternative view is provided in Appendix~\ref{fig:pipelineattacker}. In practice, this attacker model allows us to reason from observed weaknesses in firmware artifacts to their security consequences. Rather than assessing vulnerabilities in isolation, we evaluate whether they can be composed along the attack pipeline to yield effective attacker capabilities and ultimately reach one of the objectives defined in \Cref{sec:attack}. In contrast to this attacker-centric view, our analysis pipeline adopts an analyst perspective, structuring the process from firmware collection to vulnerability identification and impact assessment. This complementary pipeline is presented in Appendix~\ref{fig:our_pipeline}.

\section{Methodology and firmware corpus construction}\label{sec:methodo}

We present a methodology for the security analysis of cryptocurrency mining infrastructures that relies on the systematic examination of firmware artifacts publicly distributed by manufacturers. More precisely, we collect and analyze firmware update packages, full reflashing images, management tools, and associated technical documentation. Our approach relies exclusively on static analysis and does not require physical device acquisition, runtime interaction, or emulation. Because these artifacts correspond directly to software distributed for deployment on production devices, static analysis remains representative of real-world systems while being conducted independently of operational environments. This enables reproducible comparisons across vendors, hardware generations, and firmware versions, and makes it possible to study the mining ecosystem at scale without depending on the availability of specific devices in a laboratory setting. Our approach introduces a shift in perspective by treating publicly available firmware artifacts not as auxiliary data, but as a central and sufficient source for reconstructing device architectures, identifying vulnerabilities, and deriving complete attack paths. Mining infrastructures provide a particularly relevant case study in this context, as firmware directly governs economically sensitive parameters, including pool endpoints, wallet identifiers, frequency, voltage, and cooling. 

We aim to determine whether a device exposes the entry points, vulnerabilities, and capabilities required to reach concrete adversarial objectives. In contrast to the attacker-oriented viewpoint developed in \Cref{sec:attacker_model}, which describes how an adversary would progress from initial access to final impact, the methodology presented here adopts a global and systematic perspective. Rather than assuming a specific attack path, we enumerate devices, collect the corresponding artifacts, normalize the corpus, reconstruct analyzable firmware images, identify security-relevant weaknesses, and map them to the capabilities and objectives formalized in our attack model.

\subsection{Miner enumeration and ecosystem mapping}

We first construct an exhaustive inventory of commercially available \gls{asic} miners by aggregating data from major public mining ecosystem platforms, including AsicMinerValue \cite{AsicMinerValue}, MinerStat \cite{MinerStat} and WhatToMine \cite{WhatToMine}. The complete device inventory used for this mapping is summarized in Appendix~\ref{sec:fulllist}. This initial mapping step allows us to identify dominant vendors, hardware families and firmware lineages, and to define a consistent naming and classification scheme for subsequent analysis. We focus on the three dominant miner vendors (Bitmain, MicroBT, and Canaan) and also include Iceriver. Together, these four account for the overwhelming majority of deployed mining hardware~\cite{CambridgeMining2025}. Additionally, this synthesized list constitutes the baseline for firmware collection and analysis throughout this work. 

\begin{figure}[t]
  \centering
  \resizebox{0.50\columnwidth}{!}{%
  \begin{tikzpicture}
    \fontfamily{cmss}\selectfont
    \pie[sum=auto, text=inside, font=\large, color={bluePierre!70, redPierre, yellowPierre, greenPierre}]{123/Bitmain, 65/MicroBT, 36/Canaan, 22/Iceriver}
  \end{tikzpicture}
  }
  \caption{Distribution of commercially available \gls{asic} miners models per manufacturer, derived from aggregated public mining ecosystem platforms.}
  \label{fig:miner_list_distribution}
\end{figure}
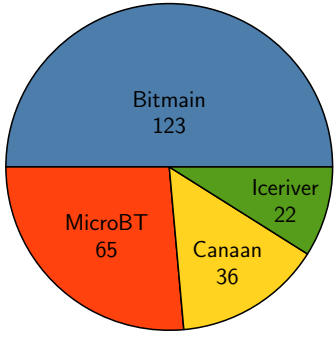

\Cref{fig:miner_list_distribution} illustrates the relative distribution of miners per manufacturer as derived from this aggregated inventory, highlighting the extreme concentration of the \gls{asic} mining ecosystem.

\subsection{Automated collection of firmware and vendor artifacts}

For each identified miner model and manufacturer, we systematically collect all publicly accessible firmware images and associated vendor artifacts, including update packages, full reflashing images, management tools and documentation. Our collection methodology deliberately relies exclusively on open and publicly reachable distribution channels, without requiring any authentication, privileged access or physical interaction with mining hardware. In practice, artifact collection follows three complementary strategies, depending on the nature of the vendor distribution infrastructure.

First, for official vendor websites exposing firmware through structured backend services, we perform automated scraping by identifying the underlying download APIs or backend endpoints used by the web interface. This typically involves enumerating API requests, catalog endpoints or JSON indices that list available firmware files. Once identified, these endpoints are queried programmatically to retrieve exhaustive file listings, which are then downloaded in bulk. This method enables massive and reproducible collection of all firmware versions made available through official vendor portals, even when the frontend interface only exposes a subset of files.

Second, when vendors distribute firmware or tooling through public version control repositories, we directly clone the corresponding repositories in their entirety. This approach preserves full version history, metadata and directory structure, and allows us to capture both current and legacy firmware releases, auxiliary scripts, configuration templates and documentation exactly as published by the vendor or its ecosystem.\cite{CanaanGitHubRepos}

\begin{figure}[bt]
  \centering
  \includegraphics[width=\columnwidth,trim={0 10.25cm 3.5cm 0},clip]{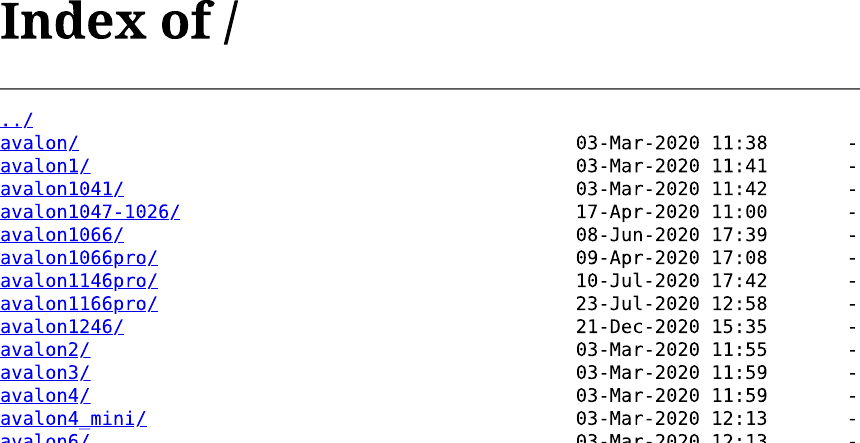}
  \caption{An excerpt from the website \url{https://download.canaan-creative.com/}}
  \label{fig:download-canaan}
\end{figure}

Third, for vendors hosting firmware artifacts on static file servers exposing directory listings, such as publicly accessible HTTP indexes, we deploy a recursive crawler to reconstruct the full directory hierarchy. This category includes sites that directly expose their internal storage layout, often backed by cloud object storage infrastructures. In these cases, the crawler enumerates directories and subdirectories, mirrors the site architecture, and selectively downloads relevant artifacts based on filename patterns, file extensions and directory context. \Cref{fig:download-canaan} illustrates a representative example of such a directory firmware repository.

\subsection{Corpus normalization and firmware reconstruction}

The collected firmware corpus is inherently heterogeneous, spanning multiple vendors, device families, time periods, and artifact types, including update packages, flash images, management tools, and documentation. To enable systematic analysis, we apply a normalization and reconstruction pipeline that transforms raw artifacts into a consistent and analyzable dataset.

Each artifact is associated with a structured identifier composed of manufacturer, miner family, and firmware generation, inferred from filenames, embedded metadata, configuration files, and scripts. This enables the reconstruction of firmware lineages and the separation of device-specific firmware from generic tooling. Artifacts are then categorized into functional classes, primarily distinguishing update packages from full flash images. Since update packages often contain only partial components, we reconstruct complete firmware images through decompression, decryption, and filesystem extraction using standard tools (e.g., \texttt{binwalk}, \texttt{ubireader}), together with custom scripts and publicly available reverse-engineering utilities for proprietary formats.

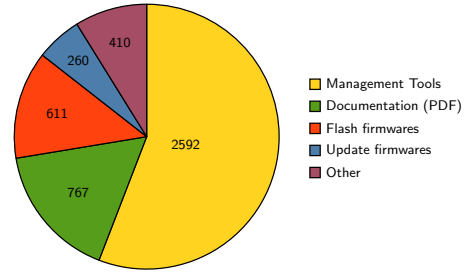
\begin{figure}[ht!]
    \centering
    \resizebox{0.70\columnwidth}{!}{
    \begin{tikzpicture}
      \fontfamily{cmss}\selectfont
      \pie[
        sum=auto,
        text=legend,
        font=\small,
        rotate=90,
        change direction,
        color={yellowPierre, greenPierre, redPierre, bluePierre!70, brownPierre!70}
      ]{
        2592/Management Tools,
        767/Documentation (PDF),
        611/Flash firmwares,
        260/Update firmwares,
        410/Other
      }
    \end{tikzpicture}
    }
    \caption{Distribution of collected firmware-related artifacts before filtering.}
    \label{fig:files_stats_appendix}
\end{figure}

The initial dataset contains 871 firmware-related artifacts (\Cref{fig:files_stats_appendix}), extracted from a broader collection of 3,769 files including documentation and tooling. Although this corpus provides broad coverage across vendors, device families, and firmware generations, it exhibits significant heterogeneity in format, completeness, and redundancy, which prevents direct large-scale analysis without preprocessing. Our objective is to maximize analytical coverage while avoiding redundant work, as analyzing multiple artifacts corresponding to equivalent firmware instances does not provide additional security insight but significantly increases analysis cost.

We first apply integrity and format validation to eliminate corrupted, truncated, or invalid artifacts that cannot be processed by standard tooling. We then attempt to decrypt, decompress, and unpack all remaining artifacts. Firmware packages frequently embed proprietary container formats, encryption schemes, or nested compression layers, which we handle using a combination of standard tools, custom scripts, and publicly available reverse-engineering utilities.

For proprietary formats, extraction often relies on external technical knowledge and community reverse engineering efforts. Bitmain firmware distributed in the \texttt{.bmu} format requires dedicated unpacking logic together with format-specific handling, which we perform using publicly available tools such as BitmainFirmwareUnpacker~\cite{BitmainFirmwareUnpacker}. Iceriver firmware similarly relies on encrypted update containers. Reverse engineering efforts, notably those associated with custom firmware projects such as \texttt{iceriver-oc}~\cite{IceRiverOC}, have shown that these containers rely on static, hardcoded secrets combined with standard symmetric cryptographic primitives. Once these secrets are recovered, the decryption process becomes reproducible and allows full extraction of embedded firmware components.

Comparable efforts exist for MicroBT and related WhatsMiner ecosystems. However, reconstruction remains incomplete for a subset of artifacts due to stronger obfuscation, partially undocumented formats, and unresolved encryption or missing cryptographic material, leading to their exclusion from further analysis.

We then evaluate reconstruction completeness. Many collected files correspond only to partial update payloads and do not expose a full system image. Since our analysis requires a coherent firmware environment, we retain only artifacts that allow reconstruction of a mountable filesystem together with executable binaries, configuration files, and sufficient contextual information to interpret update logic and exposed services. Artifacts that remain too partial to support reliable structural analysis are discarded.

The dominant reduction factor is redundancy. Firmware ecosystems exhibit extensive reuse across models, generations, and minor releases, with many artifacts differing only by packaging variations, metadata changes, or limited version increments. A large fraction of the remaining artifacts corresponds to near-identical firmware images. Retaining all of them would artificially skew the analysis by over-representing specific firmware lineages without increasing the diversity of observable security properties.

To address this, we perform large-scale deduplication using binary similarity and structural comparison techniques, clustering similar artifacts and retaining a single representative image per cluster unless a variant provides meaningful cross-version differences. This process yields a final dataset of 134 distinct firmware images, including 102 Bitmain images, 9 MicroBT images, 12 Canaan images, and 11 Iceriver images.

Overall, the reduction is driven primarily by redundancy rather than by data loss, which highlights a structural property of the ecosystem itself: firmware components and update mechanisms are heavily reused across devices. As a result, weaknesses identified in a representative subset are likely to generalize to a much larger deployment base.

\begin{table}[ht!]
\centering
\scriptsize
\begin{tabular}{lccc}
\toprule
\textbf{Filtering stage} & \textbf{Remaining} & \textbf{Removed} & \textbf{Main cause} \\
\midrule
Initial candidates & 871 & -- & Flash + update artifacts \\
Integrity filtering & 820 & 51 & Corrupted / invalid files \\
Decryption step & 690 & 130 & Encrypted / unsupported formats \\
Reconstruction step & 612 & 78 & Partial / incremental updates \\
Deduplication & 134 & 478 & Redundant variants \\
\bottomrule
\end{tabular}
\caption{Corpus reduction pipeline. Redundancy across firmware versions is the dominant reduction factor.}
\label{tab:filtering}
\end{table}

\subsection{Static analysis and vulnerability mapping}

We perform static analysis across the normalized firmware corpus, combining string extraction, file system inspection, and script/binary analysis to identify vulnerability patterns relevant to the attack model of \Cref{sec:attacker_model}. Concretely, we search for hard-coded credentials in configuration files and binaries, authentication logic in web interfaces and APIs, update scripts and signature verification routines, exposed services (e.g., SSH, HTTP servers), and embedded cryptographic material such as keys, certificates, or initialization vectors.
The analysis relies on standard tooling (e.g., \texttt{binwalk}, \texttt{strings}, file system parsers) as well as custom scripts to automatically extract and index relevant artifacts across all firmware images. For instance, we identify credential reuse through pattern matching in shadow files and initialization scripts, detect weak update mechanisms by inspecting flashing scripts and signature checks, and enumerate exposed services by analyzing startup scripts and configuration files. Network-related components are also examined to determine whether communications are authenticated or encrypted, and whether APIs enforce access control.
The objective is not to list all technical weaknesses, but to isolate those that are reachable given the entry points identified in the firmware. Automated vulnerability extraction produces a large number of candidate findings, which are then manually reviewed to confirm their validity and relevance. This step is necessary because firmware images often include unused components, legacy code, or generic tools that may appear vulnerable but are not actually exposed in the deployed system.

\subsection{Attack surface inference}

Building on the identified vulnerabilities, we determine whether they form coherent attack scenarios consistent with the model defined in \Cref{sec:attacker_model}. Rather than analyzing weaknesses in isolation, we evaluate how they combine with exposed entry points to enable concrete attacker actions.

For each firmware image, we identify the most impactful scenario that can realistically be constructed. When multiple attack paths are possible, we retain the one leading to the highest level of compromise according to the capability hierarchy. This allows us to associate each device with a dominant attack scenario reflecting its worst-case exposure.

This classification provides a simplified but operational view of the attack surface. Instead of reporting raw vulnerabilities, each miner is characterized by the strongest attack it enables, which makes it possible to compare devices and to identify broader trends across manufacturers. In particular, it allows us to determine whether a given vendor ecosystem tends to expose only limited attack paths or systematically enables high-impact compromises.

\subsection{Scope and interpretation of vulnerabilities}

Our analysis is restricted to weaknesses that can be linked to concrete attack paths, rather than attempting to exhaustively report all flaws present in the firmware.

The methodology is purely static. We do not rely on emulation or runtime validation, which prevents direct observation of dynamic behaviors such as memory corruption or runtime privilege escalation. In addition, firmware artifacts are often partial and may omit critical components such as bootloaders or hardware-specific services. As a result, some protections present on deployed devices may not be visible in the analyzed images (e.g. a firewall). These limitations affect exploit validation but not structural analysis. The objective is to determine whether the conditions required for an attack scenario are present, not to guarantee end-to-end exploitability in all configurations. To mitigate these limitations, we complement the static analysis with targeted validation on real devices, as described in Section~\ref{sec:validation}. For each of the 134 reconstructed firmware images, we manually verified the presence of the identified weaknesses and their associated attack paths whenever direct inspection was possible. This validation step ensures that the reported findings correspond to actual device behavior and not only to artifacts of static analysis.

\section{Information retrieved}\label{sec:inforetri}

The purpose of this first section is to show that, beyond security aspects, a significant amount of critical information can already be recovered from these firmware images.

\subsection{OS}
Once unpacked, update artifacts reveal concrete and verifiable properties of the deployed software stack. These include the embedded Linux family, filesystem organization, init conventions, kernel lineage, and vendor-specific components. Such information is difficult to infer from network behavior alone. This insight directly guides static analysis by narrowing the expected locations of credentials, update logic, exposed services, and management interfaces.

Out of the 134 extracted artifacts, we are able to identify the underlying \gls{os} for 121 images.  The corpus is primarily based on Angstrom or OpenEmbedded, followed by Buildroot and OpenWrt.  A small subset of images remains unclassified due to missing or non-recoverable root filesystem components.

These \gls{os} fingerprints already provide useful guidance for analysis. OpenWrt images expose stable and easily identifiable configuration structures such as \texttt{/etc/config} and \texttt{uci}, whereas Buildroot and OpenEmbedded images more often correspond to tightly controlled vendor appliances with custom layouts and limited traces of standardized package management.

Vendor distributions reflect differing engineering choices rather than uniform platform shifts. Bitmain spans multiple \gls{os} families across its update images, while MicroBT relies exclusively on OpenWrt in our dataset. Canaan exhibits a mixed profile across the Avalon family. Iceriver stands apart, as none of its update packages expose a recoverable root filesystem, which is consistent with a partial update strategy distributing only tightly packaged components.

\subsection{Miner}
Our empirical analysis highlights a strong structural convergence at the software level. Despite heterogeneity in hardware architectures and control logic, we find that the mining functionality itself consistently relies on a small set of long-established open-source projects, predominantly appearing as outdated releases or heavily customized forks.

Across the analyzed corpus, \textit{Iceriver} devices rely on \emph{cpuminer-multi} \cite{cpuminer-multi-github,cpuminer-original} (version 1.3.7), a \gls{cpu}-oriented mining software whose codebase originates from an older generation of cryptocurrency mining tools. \textit{MicroBT} devices initially embed \texttt{cgminer}~\cite{cgminer-github} version~4.9.2, before transitioning in more recent generations to a proprietary binary referred to as \texttt{btminer}. While the latter obscures its upstream origin, configuration semantics, Stratum protocol handling, and internal control patterns strongly indicate continuity with a \texttt{cgminer} 4.9.x codebase rather than a ground-up reimplementation. \textit{Canaan} firmware exhibits a broader historical span, ranging from \texttt{cgminer}~4.0.0 in older devices to \texttt{cgminer}~4.11.1, the final official upstream release, in more recent models. \textit{Bitmain} devices show a similar evolution, progressing from early \texttt{cgminer}~3.4.3 deployments to \texttt{cgminer}~4.9.x, before consolidating around a proprietary fork known as \texttt{bmminer}, which retains the architectural and protocol foundations of \texttt{cgminer} while introducing vendor-specific extensions.

Taken together, these observations show that the mining layer of contemporary miner firmware ecosystems remains anchored in software projects whose main upstream development dates back roughly to the 2014--2018 period. In other words, even in recent devices, the core mining logic often relies on software stacks designed and released several years ago, then incrementally adapted rather than fundamentally renewed. Vendors extend these legacy codebases through external management binaries, configuration layers, and hardware-specific drivers, effectively freezing the core mining logic while evolving the surrounding control infrastructure.

This widespread reliance on aging open-source software has important implications. \texttt{cgminer} is distributed under the GNU General Public License version~3\footnote{We reference upstream licenses solely to explain software lineage and security implications. We did not audit license compliance and make no legal claims.}, while \texttt{cpuminer-multi} is licensed under version~2 of the same license family. While license compliance is out of scope, this discussion clarifies software lineage and highlights the central role of open-source components in an ecosystem otherwise presented as highly proprietary.

From a security perspective, this design choice concentrates risk on a limited set of legacy codebases. For example, \texttt{cgminer} versions up to 4.10.0 contained an authenticated remote code execution vulnerability in the remote management interface (CVE-2018-10058)~\cite{CVE2018Cgminer}. Beyond known vulnerabilities, we performed a lightweight static analysis of \texttt{cgminer} and \texttt{cpuminer-multi}, two representative mining cores that remain widely reused across vendors. Our analysis relies on \emph{Semgrep}~\cite{semgrep}, complemented by manual inspection of the upstream codebases~\cite{cgminer-github,cpuminer-multi-github}.

Across both projects, we consistently observe recurring classes of weaknesses. First, multiple code paths rely on unbounded string manipulation functions without explicit bounds checking, increasing the risk of buffer overflows in the presence of malformed inputs. Second, the codebases exhibit fragile memory management patterns, with manual allocation and deallocation logic that provides limited safety guarantees and may lead to use-after-free conditions or insufficient validation of allocation results. Third, externally controlled data originating from mining pools, configuration interfaces, or remote management channels is not systematically sanitized before being processed, exposing the software to injection or parsing vulnerabilities. Finally, several control and monitoring interfaces operate over plaintext communication channels, exposing sensitive operational data and increasing the attack surface for network-based adversaries.

Given that these mining cores are reused across vendors through customized forks, often with limited visibility and reduced community review, such weaknesses may persist and propagate across independently maintained firmware distributions. As a result, the mining layer itself constitutes a shared and largely homogeneous attack surface across otherwise heterogeneous devices.

\section{Security exposures}\label{sec:security_exp}
Security weaknesses in cryptocurrency mining devices are partially documented, including exposed management interfaces, hard-coded credentials, insecure update mechanisms, and backdoors~\cite{queenant2016,antbleed2017,cve2018bitmain,cve2022avalon,cve2022goldshell1,cve2022goldshell2,cve2022goldshell3,chambers2022,dcentral2025}. However, these findings remain fragmented and lack a unified mapping to concrete attacker capabilities.

We therefore systematize existing knowledge by linking vulnerabilities to operational attack scenarios. Our analysis shows that publicly available firmware artifacts already constitute a major attack surface: without device access, an adversary can reconstruct internal architecture, identify weaknesses, and prepare attacks offline at negligible cost.

This challenges the current firmware trust model, where update artifacts implicitly enable large-scale reconnaissance. In the following, we map these weaknesses to realistic attacker actions and impacts.

We have included a more comprehensive analysis of a model by vendor in the appendix \Cref{sec:deeper_analysis}.

\subsection{Firmware phishing attack}

All major manufacturers rely on manual update workflows, in which operators are expected to download and flash firmware images obtained from vendor websites. This design inherently creates a highly scalable phishing attack surface.

Across all major manufacturers, mining devices explicitly support reflashing using external firmware packages, including \gls{sd} card based procedures, and routinely accept firmware originating outside official distribution infrastructures:
\begin{itemize}
    \item \textbf{Bitmain} documents \gls{sd} card flashing and supports external firmware deployment \cite{BitmainSDFlash,AwesomeMinerBitmain}.
    \item \textbf{MicroBT} supports \gls{sd} card flashing and has a large third-party firmware ecosystem \cite{ZeusBTCWhatsMiner,BixbitWhatsMiner,WhatsMinerYT,HitsxxWhatsminerUpgrade}.
    \item \textbf{Canaan} firmware images are publicly redistributed through multiple external reseller platforms \cite{ZeusBTCAvalon}.
    \item \textbf{Iceriver} firmware packages are similarly redistributed by third-party vendors \cite{ZeusBTCIceRiver}.
\end{itemize}

As long as a firmware package conforms to the expected update container format and passes basic validation logic, it gains persistent privileged execution on the device.
This enables a class of social engineering attacks that we term \emph{firmware phishing attacks}, in which adversaries distribute malicious firmware images masquerading as legitimate updates. Such images can embed wallet redirection logic, hidden mining pools, persistent backdoors, or destructive overclocking parameters.

\begin{figure}[bt]
  \centering
  \includegraphics[width=\columnwidth,trim={30mm 15mm 32mm 0},clip]{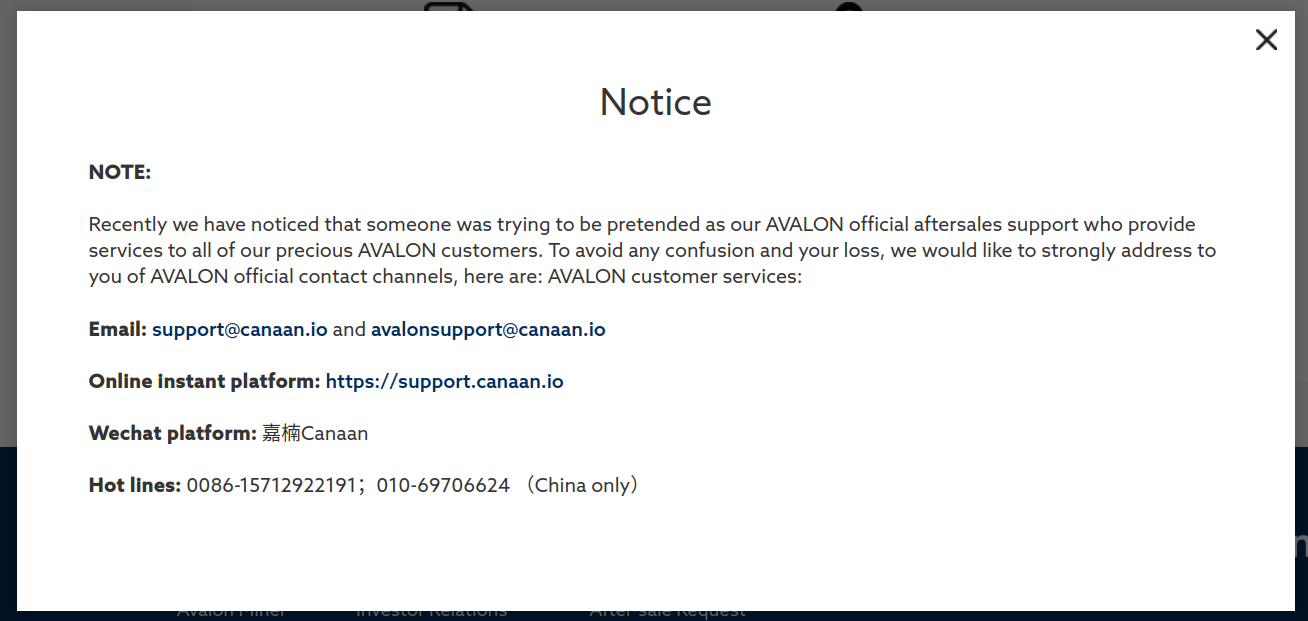}
  \caption{Warning displayed on the official Canaan support website in 2025, alerting users about fraudulent firmware update instructions and impersonation of support channels.}
  \label{fig:warningcanaan}
\end{figure}

These attacks can be deployed at scale through impersonation of vendor after-sales support, for instance by advertising urgent security fixes or mandatory performance updates. Notably, this threat is explicitly acknowledged by manufacturers themselves. Canaan publicly warns users about ongoing impersonation of official support channels used to distribute fraudulent firmware updates \cite{CanaanNotice}. We observed this warning displayed on Canaan’s official support website in 2025, as shown in \Cref{fig:warningcanaan}, indicating that firmware phishing is an active real-world attack vector rather than a hypothetical risk.

In the absence of a strictly enforced secure boot mechanism with cryptographic root of trust, firmware authenticity ultimately relies on operator judgment. Firmware phishing therefore enables persistent compromise~\cite{NotepadPlusPlusHijacked2026} of mining fleets without requiring software vulnerabilities, runtime exploits, or physical access, constituting a systemic weakness of the \gls{asic} mining ecosystem. This directly enables the attack scenario of \emph{full device takeover and infrastructure abuse}.

\subsection{Systemic use of stratum V1 over plaintext TCP}

We analyzed mining binaries, configuration files, startup scripts and vendor documentation across the full firmware corpus to identify supported mining communication protocols. Across all manufacturers, we did not observe any native support for Stratum~V2 (a successor protocol with authenticated and encrypted transport) or mandatory encrypted Stratum channels. 
All devices rely on Stratum~V1 and by default operate over plaintext TCP connections.

Stratum~V1 was designed for simplicity and low overhead, but provides no transport encryption, no authentication of pool endpoints, and no cryptographic integrity for job assignments or share submissions. Consequently, all mining pool communication is transmitted in cleartext and is fully malleable in transit. This design creates a systemic man-in-the-middle attack surface. Any adversary positioned on the network path, including local attackers, compromised gateways, or \gls{isp}s, can intercept, modify, and replay Stratum traffic. By altering wallet addresses, pool endpoints, or job parameters, an attacker can silently redirect hashrate and rewards without disrupting mining operations.

These weaknesses are well documented in prior work. Stratum~V1 has been shown to enable stealth hashrate theft, share manipulation, and proxy redirection attacks~\cite{liu2021disappeared,recabarren2017hardeningstratum,StratumMITM,StratumHijack,StratumAttackSurface}. Crucially, the vulnerability resides at the protocol layer itself and therefore affects miners regardless of firmware hardening or web interface security.

Vendor documentation further reinforces this exposure. For most manufacturers, Stratum over plaintext TCP is the default and commonly documented configuration, with no requirement to enable encryption. Only Iceriver and Canaan document optional support for SSL or TLS endpoints, while still allowing unencrypted TCP connections. As a result, even when encrypted Stratum is technically available, nothing encourage user from deploying miners in an insecure plaintext mode. This makes miners vulnerable to the \emph{Cryptocurrency theft and revenue redirection} attack scenario.

\subsection{Local network threat model and high impact defaults}

This section does not aim to exhaustively enumerate individual vulnerabilities. Instead, it focuses on identifying which attack scenarios (as defined in \Cref{sec:attack}) are realistically achievable for each manufacturer, based on the exposed attack surface and observed configurations.

The vulnerabilities discussed in the following subsections primarily matter for an adversary who can reach miners on their management network (\gls{lan} or internal fleet segment). In practice, miners are rarely exposed directly to the public Internet.\footnote{To the best of our knowledge, an analysis conducted using \url{https://www.shodan.io/} did not reveal any mining devices directly exposed to the public Internet.} They are typically placed behind \gls{nat}s, firewalls, or dedicated \gls{vlan}s and administered from nearby workstations or management servers~\cite{enisa2019_threat_landscape_2018}.This prerequisite commonly holds in mining and hosting environments, where attackers may gain internal access through compromised infrastructure, misconfigurations, or lateral movement. Under this model, weak credentials combined with reachable services (\gls{ssh}, web UI, or management APIs) provide immediate high-impact primitives, including persistent takeover, configuration tampering, and hashrate redirection.

This type of attack remains realistic and well-documented in practice, as illustrated by prior work on lateral movement and internal network compromise, as well as real-world incidents affecting large-scale infrastructures~\cite{274594,247682,11022024,NYT_MGM_2023,WSJ_UnitedHealth_2024}.

As summarized in Table~\ref{tab:attack_summary_models}, these attack primitives translate into widespread exploitability across vendors under a LAN threat model. The list in Appendix~\ref{sec:fulllist} also indicates which scenario we consider vulnerable to each model.

\subsubsection{Bitmain}

For Bitmain devices, we adopt a near exhaustive per model analysis strategy.  Our firmware corpus contains almost one image per miner model or hardware family (102 usable images in total), covering 113 out of 123 Bitmain miners, excluding only the S23, KS7 and L11 families for which no extractable artifacts were available. Across all analyzed Bitmain firmware images, we consistently identify the presence of at least one user account protected by a weak or default password together with an active \gls{ssh} service. This configuration alone is sufficient to enable immediate remote administrative access, without requiring exploitation of software vulnerabilities. As a result, all analyzed Bitmain devices are directly susceptible to our strongest attack class, namely \emph{full device takeover and infrastructure abuse} (see \Cref{sec:attack}). Although not all S21 miners ship with active local user accounts, the factory \texttt{/etc/shadow.factory} file still embeds a default password and SSH remains enabled, while the presence of an unauthenticated cleartext HTTP management interface and API remains exploitable, enabling at minimum \emph{Cryptocurrency theft and revenue redirection}. 

%

\subsubsection{MicroBT} 

For MicroBT devices, firmware update artifacts could not be reliably extracted despite extensive static analysis efforts.  Publicly distributed update packages exhibit high entropy and strong obfuscation, preventing reconstruction of usable update firmware images. As a result, our analysis relies on the alternative recovery mechanism supported by MicroBT~\cite{WhatsMinerYT}. These flashing images provide only partial visibility into the software stack: the mining application and web management interface are absent, and significant portions of the image are encrypted.  In addition, miners still require a subsequent online firmware update after reflashing, further limiting the completeness of the recovered material.

Unlike Bitmain, MicroBT firmware organization follows a control board centric model.  We therefore structured our analysis by associating firmware artifacts with control board families (H616, H6OS, H6, and H3) rather than individual miner models. Under these constraints, our analysis does not expose direct administrative takeover primitives. 
The only demonstrable weakness corresponds to \emph{Performance degradation and operational disruption} attack scenario, where network reachable management interfaces during discovery phases allow interactions that can disrupt normal miner operation.
%

\subsubsection{Canaan}

For Canaan devices, we successfully collected and extracted firmware images spanning a broad historical and product range, from early Avalon generations to recent consumer and professional models. Firmware and update images for Avalon~1 through Avalon~9, as well as Avalon~15 and the Avalon family, covering 18 out of 36 miners, were obtained from publicly accessible official sources, including the vendor download portal and public source code repositories (\Cref{fig:download-canaan}) \cite{CanaanFirmwarePortal,CanaanGitHubRepos}. These resources provide substantial visibility into firmware versions, update workflows, tooling, and documentation, enabling comprehensive offline analysis without requiring physical hardware. All fully extracted Avalon firmware images are vulnerable to the strongest attack scenario identified in this work, namely \emph{full device takeover and infrastructure abuse}, due to an exposed \gls{ssh} service combined with a factory account protected by weak credentials. For Avalon models where complete firmware reconstruction was not possible, complementary evidence from public vendor resources indicates exposure to a secondary yet critical attack scenario: \emph{Cryptocurrency theft and revenue redirection}. In these cases, the management API follows an overly permissive trust model, granting administrative privileges to any authenticated client. This allows attackers with network access to modify mining pool endpoints, payout addresses, and hardware settings. Authentication exchanges for sensitive actions are performed over plaintext channels, making them vulnerable to interception or replay by on-path adversaries. For certain generations (notably Avalon~12 to Avalon~14), vendor-documented API commands further allow the administrative password to be reset to a default value, effectively removing any remaining protection. Beyond revenue redirection, the exposed API also enables repeated fan and power control operations, potentially leading to performance degradation or hardware damage.

\subsubsection{Iceriver}

In contrast with other manufacturers, Iceriver firmware update packages are structurally partial and typically expose only isolated components, such as application binaries or web interface resources, rather than a full root filesystem. This limits the amount of information recoverable through static analysis and prevents reliable identification of the underlying operating system or default services such as \gls{ssh}.

Despite this limited visibility, we identified a critical weakness in the authentication logic of the Iceriver web management interface, which is shared across multiple miner models and generations. This flaw allows unauthorized access to administrative functions and enables direct modification of mining pool endpoints and wallet addresses. As a result, Iceriver devices are vulnerable to the \emph{Cryptocurrency theft and revenue redirection} attack scenario described in \Cref{sec:attack}, even in the absence of full firmware visibility.

\begin{table}[t]
\centering
\scriptsize
\setlength{\tabcolsep}{3pt}
\renewcommand{\arraystretch}{1.1}
\begin{tabular}{lcccc}
\toprule
\textbf{Attack scenario on LAN} 
& \textbf{Bitmain (123)} 
& \textbf{MicroBT (65)} 
& \textbf{Canaan (36)} 
& \textbf{Iceriver (22)} \\
\midrule

Full device takeover 
& 113 
& 0 
& 26 
& 0 \\

Revenue redirection 
& 113 
& 0 
& 34
& 22 \\

Physical degradation 
& 113 
& 0 
& 34
& 22 \\

Performance disruption 
& 113 
& 65 
& 34
& 22 \\

\bottomrule
\end{tabular}
\caption{LAN attack scenarios: number of vulnerable miner models per manufacturer (totals in parentheses). Further details in \Cref{sec:fulllist}.}
\label{tab:attack_summary_models}
\end{table}
\section{Validation on real devices}\label{sec:validation}

To validate the relevance of our firmware findings, we complemented our analysis with measurements on two real devices. For each miner, we extracted data through two complementary paths: live acquisition by leveraging identified weaknesses to access and collect runtime artifacts, and cold NAND flash extraction to recover the most complete filesystem and boot material possible.  In addition, we experimentally evaluated the attack scenarios introduced in Section~\ref{sec:attack} to assess their feasibility on real hardware. We applied this methodology to an Iceriver AL0 miner and to a Bitmain Antminer KS5, enabling a direct comparison between publicly distributed artifacts and the full firmware state.

\paragraph{Iceriver.}Although Iceriver update packages are often structurally partial, our cross-artifact comparison shows that the security-relevant components they do expose are consistent with those observed on a real device. In particular, the embedded web management stack and boot artifacts such as \texttt{BOOT.BIN} closely match what we extracted from an Iceriver AL0 miner in our lab.  Notably, no official firmware update package is publicly available for the AL0 model. Nevertheless, the consistency observed across artifacts strongly suggests that Iceriver reuses the same firmware building blocks across models and generations. In particular, our boot analysis indicates reuse of the same Zynq boot chain (FSBL and U-Boot~2019.01) across multiple \texttt{BOOT.BIN} variants, which strongly supports the hypothesis of a shared firmware lineage across devices and generations. This reuse suggests that vulnerabilities identified in publicly distributed, partial update artifacts can plausibly generalize to real-world deployments, as manufacturers tend to repackage and redeploy the same firmware building blocks. Importantly, analysis of the real AL0 filesystem reveals the presence of a preconfigured local user account (\texttt{Miner168861}) with a persistent password hash provisioned by the manufacturer, together with an active \gls{ssh} service. This condition could not be observed from vendor-distributed artifacts alone, both because these update packages do not expose a complete root filesystem and because no official update artifact is available for the AL0 miner model and indicates that deployed systems may expose additional attack surfaces beyond what update packages suggest. Finally, the client-side weakness identified in \texttt{login.js} (cookie handling) is present in both the downloaded web interface and the version extracted from the real device, reinforcing that this issue is not an artifact of incomplete update packages but a persistent flaw across Iceriver deployments. This configuration directly enables the \emph{Full device takeover and infrastructure abuse} attack scenario defined in Section~\ref{sec:attack}, a risk that remains largely invisible when considering firmware artifacts alone and was therefore underestimated in the previous analysis.

\paragraph{Bitmain.}For the Bitmain Antminer KS5, we performed a direct file comparison between the publicly distributed firmware image (release update payload) and the full filesystem extracted from the device eMMC. The comparison reveals that the delta between the two artifacts is limited and largely concentrated in configuration values and recompilation timestamps rather than in structural security hardening. In particular, we observe changes in the factory mining configuration (e.g., frequency and voltage defaults in \texttt{/etc/cgminer.conf.factory}) and in thermal management parameters (e.g., a non-zero minimum PWM threshold in \texttt{/etc/topol.conf}), which primarily affect performance and operational stability. We also identify minor differences in the update CGI scripts of the web interface, such as removing a default fallback value when emitting the \texttt{bitmain-work-mode} field and enabling \texttt{dmesg} output in a log endpoint. In addition, several core binaries (\texttt{busybox}, \texttt{lighttpd}, \texttt{ntpd}, and the mining control binary \texttt{godminer}) differ by hash and embed distinct build dates, indicating that the device image and the release artifact were built at different times or from slightly different source snapshots. However, from a security standpoint, these changes do not materially alter the dominant risk profile: the real device still exposes remote administration surfaces (notably \gls{ssh}) combined with weak credential assumptions, which remains sufficient to enable full compromise by any attacker with network access. This KS5 validation therefore supports a key conclusion of our work: while vendors may introduce incremental updates and operational tuning between distributed artifacts and deployed images, the persistence of high impact access primitives (e.g., password remote shells) dominates the security posture and renders smaller hardening deltas largely irrelevant in realistic threat models.

\section{Security recommendations}\label{sec:sugg}

Our results show that the security of \gls{asic} mining infrastructures is determined primarily by the integrity of the firmware lifecycle rather than by isolated flaws. Mitigations should therefore focus on restoring clear trust boundaries across firmware distribution, installation, and execution.

More fundamentally, these measures aim to disrupt the attacker pipeline introduced in Figure~\ref{fig:pipelineattacker}. In particular, they seek to break the progression from access to impact by reducing exposed entry points and eliminating exploitable vulnerabilities, thereby limiting the number of viable attack paths in the model of Figure~\ref{fig:attack_model}. Concretely, this includes preventing trivial access to firmware artifacts, hindering their extraction or decryption, and mitigating their misuse for malicious updates or device takeover.

This section outlines a set of implementation-level directions derived from the failure modes identified in Section~\ref{sec:security_exp}. Rather than providing exhaustive guidelines, the objective is to characterize the minimal security properties required to restore well-defined trust boundaries across firmware execution, update, and management planes.

A primary requirement is the establishment of a root of trust anchored in hardware, ensuring that only authenticated firmware can execute on the device. In practice, this relies on a secure boot chain in which each stage verifies the integrity and authenticity of the next using public key signatures. The corresponding verification key must be embedded in immutable storage, and verification must occur prior to execution. In addition, version control mechanisms are necessary to prevent rollback to vulnerable firmware versions, thereby enforcing forward-only evolution of the software stack~\cite{NIST800193}.

Firmware update mechanisms must guarantee authenticity, integrity, and freshness. This implies that update packages are systematically authenticated before installation and that version monotonicity is enforced through trusted counters or equivalent mechanisms. A clear separation between verification and installation logic reduces attack surface, while update procedures should be designed to remain atomic, with explicit recovery paths in case of failure. These properties are consistent with the design principles of secure update frameworks such as TUF and Uptane~\cite{Kuppusamy2017TUF,Uptane2019}. In particular, strengthening update mechanisms directly prevents attackers from leveraging firmware access to achieve persistence or large-scale compromise.

The public availability of firmware artifacts requires that distribution channels provide authenticity guarantees independent of operator trust. This can be achieved through signed manifests, reproducible artifact identification via cryptographic hashes, and authenticated transport mechanisms. Such measures reduce exposure to firmware substitution and phishing attacks, and align with established supply chain security practices~\cite{ENISA2019SupplyChain}.

Given the extensive code reuse across devices, consistent maintenance of shared components is required, supported by coordinated patching strategies. The extensive reuse of legacy components across firmware images implies that vulnerabilities propagate across product lines and persist over time. This necessitates a coordinated maintenance strategy at the vendor level, including tracking of embedded dependencies and synchronized patch deployment. The use of Software Bills of Materials facilitates this process by providing visibility into third-party components and associated risks~\cite{NTIASSBOM2021}.

Our empirical analysis shows that many high-impact attack scenarios stem from weak default configurations rather than complex software vulnerabilities. As a result, baseline security critically depends on secure-by-default settings, including the absence of default credentials, the restriction of remote access interfaces, and the use of strong authentication mechanisms. In particular, password-based remote access introduces unnecessary exposure and should be avoided in favor of stronger authentication schemes, in line with IoT baseline security standards~\cite{ETSI303645}. These measures directly reduce the set of reachable entry points in the attack model.

The mining communication layer constitutes a direct economic control channel and therefore requires integrity and authenticity guarantees. The widespread use of plaintext Stratum V1 introduces a systemic attack surface at the network level. Mitigations include the use of encrypted transport channels and the adoption of protocols providing authentication and integrity protection, such as Stratum V2. Prior work has demonstrated the feasibility of traffic manipulation attacks in the absence of such protections~\cite{recabarren2017hardeningstratum,StratumMITM}.

Finally, baseline security also depends on minimal observability and response capabilities. The absence of visibility into security-relevant events limits detection and response capabilities, which motivates the integration of minimal audit mechanisms such as logging of update operations and authentication events, with integrity protection where feasible~\cite{NIST80092}.

\section{Remediation and information for manufacturers}\label{remediation}
Note for reviewers: we are in contact with the suppliers and will update this section once they have made further progress.

\section{Conclusion}\label{sec:conclusion}

Publicly distributed firmware for ASIC miners constitutes a primary attack surface at the scale of the ecosystem. By analyzing firmware from vendors covering approximately \textbf{99\%} of deployed mining hardware, we show that critical security failures are not isolated bugs but inherent properties of the firmware distribution model.

The paper establishes the following:

\begin{itemize}
    \item \textbf{The firmware update workflow is a systemic vulnerability:} update artifacts alone expose the attack surface, enable large-scale firmware phishing, and allow persistent full compromise without requiring software exploitation.
    
    \item We propose a \textbf{framework} and a \textbf{methodology} for firmware analysis, enabling reproducible and large-scale inspection of firmware artifacts without requiring physical devices, and allowing recovery of internal architectures and reconstruction of attack paths.
    
    \item We demonstrate \textbf{the insecurity of the mining ecosystem:} applying this methodology reveals that critical vulnerabilities are pervasive across all major vendors, with many devices exploitable under a LAN threat model, plaintext Stratum~V1 enabling revenue redirection, and firmware phishing extending this risk to all deployments, requiring \textbf{immediate remediation}.
\end{itemize}

This paper shows that the firmware distribution model inherently enables large-scale analysis, attack reconstruction, and cross-vendor exploitation, making security failures systemic rather than incidental.

\clearpage


\section{Ethical Considerations}
\label{app:ethics}

The primary contribution of this work is a reproducible methodology for large-scale security analysis of firmware distribution ecosystems. This methodology is applied to the case of \gls{asic} cryptocurrency miners using static analysis of firmware distribution artifacts that are publicly accessible on the Internet, with the objective of documenting systemic weaknesses in firmware lifecycles and management planes and providing mitigation guidance.

The primary stakeholders are manufacturers, operators, and hosting providers, as well as the broader cryptocurrency ecosystem whose security depends on miner availability and integrity. Publication of security findings can increase adversarial awareness. This risk is mitigated by focusing on high level failure modes and by avoiding disclosure of exploitation steps or weaponizable artifacts.

The study does not involve probing deployed miners, attempting exploitation, or interacting with production systems. All observations derive from artifacts collected via open distribution channels, including update packages, reflashing images, management tools, and documentation. Limited validation was performed on two devices under our control in a controlled environment solely to confirm that publicly distributed artifacts are representative of deployed software.

Where high impact issues were identified, we initiated vendor notification and followed a coordinated disclosure mindset. We deliberately refrain from publishing information that would materially lower the cost of exploitation, such as ready-to-use payloads, credentials, exact bypass sequences, or end-to-end attack recipes.

Our ethical analysis is guided by the principles of beneficence, respect for persons, justice, and respect for law and public interest. Publication is ethically appropriate because this ecosystem is highly concentrated and exhibits extensive code reuse across models and generations, allowing systematic weaknesses to propagate widely and persist. Security-by-obscurity around firmware distribution does not provide reliable protection and can hinder independent assessment by operators. By clarifying trust boundaries and recommending mitigations such as signed updates with anti-downgrade, safer defaults, hardened management planes, and secure mining communication, this work aims to reduce real-world harm and improve ecosystem resilience.

\bibliographystyle{IEEEtran}
\bibliography{bibliography}

@inproceedings{CostinUsenix14,
  author    = {Andrei Costin and Jonas Zaddach and Aur{\'e}lien Francillon and Davide Balzarotti},
  title     = {A {Large-Scale} Analysis of the Security of Embedded Firmwares},
  booktitle = {23rd USENIX Security Symposium (USENIX Security 14)},
  year      = {2014},
  isbn      = {978-1-931971-15-7},
  address   = {San Diego, CA, USA},
  pages     = {95--110},
  url       = {https://www.usenix.org/conference/usenixsecurity14/technical-sessions/presentation/costin},
  publisher = {USENIX Association},
  month     = aug
}

@inproceedings{costin2016dynamic,
  author    = {Andrei Costin and Apostolis Zarras and Aur{\'e}lien Francillon},
  title     = {Automated Dynamic Firmware Analysis at Scale: A Case Study on Embedded Web Interfaces},
  booktitle = {Proceedings of the 11th ACM Asia Conference on Computer and Communications Security},
  year      = {2016},
  pages     = {437--448},
  publisher = {Association for Computing Machinery},
  address   = {New York, NY, USA},
  doi       = {10.1145/2897845.2897900}
}

@inproceedings{BaFIRMAFLUsenix19,
  author    = {Yaowen Zheng and Ali Davanian and Heng Yin and Chengyu Song and Hongsong Zhu and Limin Sun},
  title     = {{FIRM-AFL}: {High-Throughput} Greybox Fuzzing of {IoT} Firmware via Augmented Process Emulation},
  booktitle = {28th USENIX Security Symposium (USENIX Security 19)},
  year      = {2019},
  isbn      = {978-1-939133-06-9},
  address   = {Santa Clara, CA, USA},
  pages     = {1099--1114},
  url       = {https://www.usenix.org/conference/usenixsecurity19/presentation/zheng},
  publisher = {USENIX Association},
  month     = aug
}

@article{ZHANG2024106114,
  author   = {Xiangyu Zhang and Zhuming Chen and Shengyu Wang},
  title    = {A Study of the Impact of Cryptocurrency Price Volatility on the Stock and Gold Markets},
  journal  = {Finance Research Letters},
  volume   = {69},
  pages    = {106114},
  year     = {2024},
  issn     = {1544-6123},
  doi      = {10.1016/j.frl.2024.106114},
  url      = {https://www.sciencedirect.com/science/article/pii/S1544612324011437}
}

@misc{bao2025bibliometricanalysisscientificpublications,
  author        = {Lingfeng Bao and Jiameng Yang and Xiaohu Yang and Chunming Rong},
  title         = {Bibliometric Analysis of Scientific Publications on Blockchain Research and Applications},
  year          = {2025},
  eprint        = {2504.13387},
  archivePrefix = {arXiv},
  primaryClass  = {cs.DL},
  url           = {https://arxiv.org/abs/2504.13387}
}

@inproceedings{FengP2IMUsenix20,
  author    = {Bo Feng and Alejandro Mera and Long Lu},
  title     = {{P2IM}: Scalable and Hardware-independent Firmware Testing via Automatic Peripheral Interface Modeling},
  booktitle = {29th USENIX Security Symposium (USENIX Security 20)},
  year      = {2020},
  isbn      = {978-1-939133-17-5},
  address   = {Virtual Event, USA},
  pages     = {1237--1254},
  url       = {https://www.usenix.org/conference/usenixsecurity20/presentation/feng},
  publisher = {USENIX Association},
  month     = aug
}

@article{ulhaq2023firmware,
  author  = {Shahid Ul Haq and Yashwant Singh and Amit Sharma and Rahul Gupta and Dipak Gupta},
  title   = {A Survey on IoT and Embedded Device Firmware Security: Architecture, Extraction Techniques, and Vulnerability Analysis Frameworks},
  journal = {Discover Internet of Things},
  year    = {2023},
  volume  = {3},
  number  = {1},
  pages   = {17},
  doi     = {10.1007/s43926-023-00045-2},
  issn    = {2730-7239},
  url     = {https://doi.org/10.1007/s43926-023-00045-2}
}

@misc{nakamoto2008bitcoin,
  author       = {Satoshi Nakamoto},
  title        = {Bitcoin: A Peer-to-Peer Electronic Cash System},
  year         = {2008},
  howpublished = {\url{https://bitcoin.org/bitcoin.pdf}},
  note         = {White paper}
}

@inproceedings{bonneau2015sok,
  author    = {Joseph Bonneau and Andrew Miller and Jeremy Clark and Arvind Narayanan and Joshua A. Kroll and Edward W. Felten},
  title     = {{SoK}: Research Perspectives and Challenges for Bitcoin and Cryptocurrencies},
  booktitle = {2015 IEEE Symposium on Security and Privacy},
  year      = {2015},
  pages     = {104--121},
  publisher = {IEEE},
  address   = {Los Alamitos, CA, USA},
  doi       = {10.1109/SP.2015.14}
}

@misc{CambridgeMining2025,
  author       = {{Cambridge Centre for Alternative Finance}},
  title        = {Cambridge Digital Mining Industry Report: Global Operations, Sentiment, and Energy Use},
  year         = {2025},
  month        = apr,
  howpublished = {\url{https://www.jbs.cam.ac.uk/wp-content/uploads/2025/04/2025-04-cambridge-digital-mining-industry-report.pdf}},
  note         = {University of Cambridge, Judge Business School, First Edition}
}

@misc{Buterin2014Ethereum,
  author       = {Vitalik Buterin},
  title        = {A Next-Generation Smart Contract and Decentralized Application Platform},
  year         = {2014},
  howpublished = {\url{https://ethereum.org/en/whitepaper/}}
}

@misc{CVE2018Cgminer,
  author       = {{MITRE Corporation}},
  title        = {CVE-2018-10058: cgminer and bfgminer Remote Management API Authenticated Code Execution},
  year         = {2018},
  howpublished = {\url{https://cve.mitre.org/cgi-bin/cvename.cgi?name=CVE-2018-10058}},
  note         = {Stack-based buffer overflow in cgminer <= 4.10.0}
}

@misc{liu2021disappeared,
  author       = {Xin Liu and Rui Chong and Yuanyuan Huang and Yingli Zhang and Qingguo Zhou},
  title        = {Disappeared Coins: Steal Hashrate in Stratum Secretly},
  year         = {2021},
  howpublished = {Black Hat Asia 2021},
  note         = {Conference presentation}
}

@inproceedings{Chaum1983BlindSignatures,
  author    = {David Chaum},
  editor    = {David Chaum and Ronald L. Rivest and Alan T. Sherman},
  title     = {Blind Signatures for Untraceable Payments},
  booktitle = {Advances in Cryptology},
  year      = {1983},
  publisher = {Springer US},
  address   = {Boston, MA, USA},
  pages     = {199--203},
  isbn      = {978-1-4757-0602-4}
}

@misc{zimba2021demystifyingcryptocurrencyminingattacks,
  author        = {Aaron Zimba and Mumbi Chishimba and Christabel Ngongola-Reinke and Tozgani Fainess Mbale},
  title         = {Demystifying Cryptocurrency Mining Attacks: A Semi-supervised Learning Approach Based on Digital Forensics and Dynamic Network Characteristics},
  year          = {2021},
  eprint        = {2102.10634},
  archivePrefix = {arXiv},
  primaryClass  = {cs.CR},
  url           = {https://arxiv.org/abs/2102.10634}
}

@misc{cgminer-github,
  author       = {Kano and contributors},
  title        = {{cgminer}: Multi-threaded Multi-pool FPGA and ASIC Miner for Bitcoin},
  year         = {2026},
  howpublished = {\url{https://github.com/kanoi/cgminer}},
  note         = {GitHub repository, fork of ckolivas/cgminer, accessed 2026-04-20}
}

@article{sari2017exploitingcryptocurrencyminers,
  author  = {Arif Sari and Seyfullah Kilic},
  title   = {Exploiting Cryptocurrency Miners with {OSINT} Techniques},
  journal = {Transactions on Networks and Communications},
  volume  = {5},
  number  = {6},
  pages   = {1--9},
  year    = {2017},
  month   = dec,
  doi     = {10.14738/tnc.56.4083}
}

@misc{BabkinFirmWar2023,
  author       = {Vladyslav Babkin},
  title        = {firmWar: An Imminent Threat to the Foundation of Computing},
  year         = {2023},
  month        = may,
  howpublished = {Black Hat Asia 2023 Briefings, Singapore},
  note         = {Presentation slides}
}

@misc{AsicMinerValue,
  key          = {ASIC Miner Value},
  title        = {{ASIC Miner Value}},
  year         = {2025},
  howpublished = {\url{https://www.asicminervalue.com}}
}

@misc{MinerStat,
  key          = {MinerStat},
  title        = {{MinerStat Mining Hardware Database}},
  year         = {2025},
  howpublished = {\url{https://minerstat.com/hardware/asics}}
}

@misc{NotepadPlusPlusHijacked2026,
  author       = {{Notepad++ Project}},
  title        = {Notepad++ Hijacked by State-Sponsored Hackers},
  year         = {2026},
  month        = feb,
  day          = {2},
  howpublished = {\url{https://notepad-plus-plus.org/news/hijacked-incident-info-update/}},
  note         = {Security incident report describing an infrastructure-level compromise of the Notepad++ update distribution channel between June and December 2025},
  urldate      = {2026-02-02}
}

@misc{NYT_MGM_2023,
  author       = {Eduardo Medina},
  title        = {{`Cybersecurity Issue'} Forces Systems Shutdown at MGM Hotels and Casinos},
  year         = {2023},
  month        = sep,
  day          = {11},
  howpublished = {\url{https://www.nytimes.com/2023/09/11/technology/mgm-cyberattack.html}},
  note         = {The New York Times}
}

@misc{WSJ_UnitedHealth_2024,
  author       = {James Rundle and Catherine Stupp},
  title        = {UnitedHealth Hack: What You Need to Know},
  year         = {2024},
  month        = may,
  day          = {2},
  howpublished = {\url{https://www.wsj.com/articles/unitedhealth-hack-what-you-need-to-know-45efc28c}},
  note         = {The Wall Street Journal}
}

@inproceedings{11022024,
  author    = {Asaduzzaman Jony and Muhammad Nazrul Islam and Rahat Ahmed Talukder},
  title     = {A Secure Token-Based Approach for DHCP Client Authentication and Replay Attack Prevention},
  booktitle = {2024 27th International Conference on Computer and Information Technology (ICCIT)},
  year      = {2024},
  pages     = {855--860},
  publisher = {IEEE},
  address   = {Los Alamitos, CA, USA},
  doi       = {10.1109/ICCIT64611.2024.11022024}
}

@inproceedings{247682,
  author    = {Amirreza Niakanlahiji and Jinpeng Wei and Md Rabbi Alam and Qingyang Wang and Bei-Tseng Chu},
  title     = {{ShadowMove}: A Stealthy Lateral Movement Strategy},
  booktitle = {29th USENIX Security Symposium (USENIX Security 20)},
  year      = {2020},
  isbn      = {978-1-939133-17-5},
  address   = {Virtual Event, USA},
  pages     = {559--576},
  url       = {https://www.usenix.org/conference/usenixsecurity20/presentation/niakanlahiji},
  publisher = {USENIX Association},
  month     = aug
}

@inproceedings{274594,
  author    = {Grant Ho and Mayank Dhiman and Devdatta Akhawe and Vern Paxson and Stefan Savage and Geoffrey M. Voelker and David Wagner},
  title     = {Hopper: Modeling and Detecting Lateral Movement},
  booktitle = {30th USENIX Security Symposium (USENIX Security 21)},
  year      = {2021},
  isbn      = {978-1-939133-24-3},
  address   = {Virtual Event, USA},
  pages     = {3093--3110},
  url       = {https://www.usenix.org/conference/usenixsecurity21/presentation/ho},
  publisher = {USENIX Association},
  month     = aug
}

@misc{WhatToMine,
  key          = {WhatToMine},
  title        = {{WhatToMine ASIC Mining Database}},
  year         = {2025},
  howpublished = {\url{https://whattomine.com/miners}}
}

@misc{BitmainSite,
  author       = {{Bitmain Technologies Ltd.}},
  title        = {Bitmain Official Website},
  year         = {2025},
  howpublished = {\url{https://www.bitmain.com}}
}

@misc{MicroBTSite,
  author       = {{MicroBT Mining}},
  title        = {WhatsMiner Official Website},
  year         = {2025},
  howpublished = {\url{https://www.whatsminer.com}}
}

@misc{CanaanSite,
  author       = {{Canaan Creative Co., Ltd.}},
  title        = {Canaan Official Website},
  year         = {2025},
  howpublished = {\url{https://www.canaan.io}}
}

@misc{Binwalk,
  author       = {Craig Heffner},
  title        = {Binwalk: Firmware Analysis Tool},
  year         = {2024},
  howpublished = {\url{https://github.com/ReFirmLabs/binwalk}}
}

@article{bakhshi2024iotfirmwarevulnerabilities,
  author  = {Taimur Bakhshi and Bogdan Ghita and Ievgeniia Kuzminykh},
  title   = {A Review of IoT Firmware Vulnerabilities and Auditing Techniques},
  journal = {Sensors},
  volume  = {24},
  number  = {2},
  pages   = {708},
  year    = {2024},
  month   = jan,
  doi     = {10.3390/s24020708}
}

@misc{UBIReader,
  author       = {Artem Shevchenko},
  title        = {UBIReader},
  year         = {2024},
  howpublished = {\url{https://github.com/jrspruitt/ubi_reader}}
}

@misc{BitmainFirmwareUnpacker,
  author       = {VladTheJunior},
  title        = {BitmainFirmwareUnpacker},
  year         = {2025},
  howpublished = {\url{https://github.com/VladTheJunior/BitmainFirmwareUnpacker}},
  note         = {Community tool for unpacking Bitmain proprietary .bmu firmware images}
}

@misc{IceRiverOC,
  author       = {rdugan},
  title        = {iceriver-oc: IceRiver Overclocking Firmware},
  year         = {2024},
  howpublished = {\url{https://github.com/rdugan/iceriver-oc}}
}

@misc{BitmainSDFlash,
  author       = {{Bitmain Technologies Ltd.}},
  title        = {S19 XP Flashing SD Card Instruction},
  year         = {2024},
  howpublished = {\url{https://support.bitmain.com/hc/en-us/articles/10202973537177-S19-XP-Flashing-SD-card-Instruction}}
}

@misc{AwesomeMinerBitmain,
  key          = {Antminer},
  title        = {Antminer S19/S21 Firmware Installation},
  year         = {2024},
  howpublished = {\url{https://support.awesomeminer.com/support/solutions/articles/35000189959-awesome-miner-antminer-s19-s21-firmware-installation}}
}

@misc{ZeusBTCWhatsMiner,
  author       = {{Zeus Mining International Co., Ltd.}},
  title        = {WhatsMiner SD Card Flashing Program},
  year         = {2024},
  howpublished = {\url{https://www.zeusbtc.com/firmware-download/details/4709-whatsminer-sd-card-flashing-program-download}}
}

@misc{CanaanFirmwarePortal,
  author       = {{Canaan Creative}},
  title        = {Canaan Creative Official Firmware Download Portal},
  year         = {2020},
  howpublished = {\url{https://download.canaan-creative.com/}},
  note         = {Accessed: 2026-01-28}
}

@misc{CanaanGitHubRepos,
  author       = {{Canaan Creative}},
  title        = {Canaan Creative Public Source Code Repositories},
  year         = {2020},
  howpublished = {\url{https://github.com/orgs/Canaan-Creative/repositories}},
  note         = {Accessed: 2026-01-28}
}

@misc{BixbitWhatsMiner,
  key          = {WhatsMiner Series Firmware},
  title        = {{WhatsMiner Series Firmware}},
  year         = {2024},
  howpublished = {\url{https://bixbit.io/en/firmwares/whatsminer-series-m2x}}
}

@misc{WhatsMinerYT,
  key          = {WhatsMiner SD Card Flashing Tutorial},
  title        = {{WhatsMiner SD Card Flashing Tutorial}},
  year         = {2024},
  howpublished = {\url{https://www.youtube.com/watch?v=WxqchyZvQkA}}
}

@misc{ZeusBTCAvalon,
  author       = {{Zeus Mining International Co., Ltd.}},
  title        = {AvalonMiner Firmware},
  year         = {2024},
  howpublished = {\url{https://www.zeusbtc.com/firmware-download/avalonminer-firmware/}}
}

@misc{ZeusBTCIceRiver,
  author       = {{Zeus Mining International Co., Ltd.}},
  title        = {IceRiver Miner Firmware},
  year         = {2024},
  howpublished = {\url{https://www.zeusbtc.com/firmware-download/iceriver-miner-firmware/}}
}

@misc{CanaanNotice,
  key          = {Canaan Official Notice on Impersonated Support},
  title        = {{Canaan Official Notice on Impersonated Support}},
  year         = {2025},
  howpublished = {\url{https://www.canaan.io/support/}}
}

@misc{cpuminer-multi-github,
  author       = {{T. Pruvot and contributors}},
  title        = {cpuminer-multi: Multi-threaded CPU Miner},
  year         = {2026},
  howpublished = {\url{https://github.com/tpruvot/cpuminer-multi}},
  note         = {GPLv2-licensed GitHub repository}
}

@misc{cpuminer-original,
  author       = {{Lucas Jones}},
  title        = {CPUMiner},
  year         = {2014},
  howpublished = {\url{https://github.com/lucasjones/cpuminer-multi}},
  note         = {Original CPUMiner fork}
}

@misc{semgrep,
  author       = {{Semgrep, Inc.}},
  title        = {Semgrep: Lightweight Static Analysis for Many Languages},
  year         = {2026},
  howpublished = {\url{https://github.com/semgrep/semgrep}},
  note         = {Version 1.150.0}
}

@misc{StratumMITM,
  author       = {Arthur Gervais and Ghassan Karame and Karl W{\"u}st and Vasilios Glykantzis},
  title        = {On the Security and Performance of Proof of Work Blockchains},
  year         = {2016},
  howpublished = {Financial Cryptography and Data Security},
  note         = {Used as a reference for Stratum-related attack surface discussion}
}

@misc{StratumAttackSurface,
  author       = {Loi Luu and others},
  title        = {The Unbearable Lightness of Bitcoin Mining},
  year         = {2015},
  howpublished = {ACM CCS},
  note         = {Reference used for mining-pool attack surface discussion}
}

@misc{StratumHijack,
  author       = {R. Konoth and others},
  title        = {Bad Leverage: A Study of Cryptocurrency Mining Malware},
  year         = {2018},
  howpublished = {IEEE Security \& Privacy},
  note         = {Reference used for malware and mining abuse discussion}
}

@techreport{wang2021asic,
  author      = {Yiqiu Sun and Haichao Yang and Wentao Zhang and Yufeng Gu},
  title       = {ASIC Design for Bitcoin Mining},
  institution = {University of Michigan},
  year        = {2021},
  note        = {EECS 570 Final Report},
  url         = {https://zwtaoumich.github.io/paper/EECS570_Final_Report.pdf}
}

@inproceedings{10.1145/3581971.3581978,
  author    = {Man-Ching Yuen and Ka-Ming Lau and Chi-Wai Yung and Ka-Fai Ng},
  title     = {Adaptive Overclocking Mining Algorithm Selection Approach in the Cryptocurrency Mining Pool},
  year      = {2023},
  isbn      = {9781450397575},
  publisher = {Association for Computing Machinery},
  address   = {New York, NY, USA},
  url       = {https://doi.org/10.1145/3581971.3581978},
  doi       = {10.1145/3581971.3581978},
  booktitle = {Proceedings of the 2022 5th International Conference on Blockchain Technology and Applications},
  pages     = {50--56},
  location  = {Xi'an, China},
  series    = {ICBTA '22}
}

@inproceedings{Tran2024RoutingAttacksMiningPools,
  author    = {Muoi Tran and Theo von Arx and Laurent Vanbever},
  title     = {Routing Attacks on Cryptocurrency Mining Pools},
  booktitle = {2024 IEEE Symposium on Security and Privacy (SP)},
  year      = {2024},
  pages     = {3805--3821},
  publisher = {IEEE},
  address   = {Los Alamitos, CA, USA},
  doi       = {10.1109/SP54263.2024.00254},
  url       = {https://doi.org/10.1109/SP54263.2024.00254}
}

@misc{recabarren2017hardeningstratum,
  author        = {Ruben Recabarren and Bogdan Carbunar},
  title         = {Hardening Stratum, the Bitcoin Pool Mining Protocol},
  year          = {2017},
  eprint        = {1703.06545},
  archivePrefix = {arXiv},
  primaryClass  = {cs.CR},
  url           = {https://arxiv.org/abs/1703.06545}
}

@misc{merces2018trendmicro_iot_miner,
  author       = {Fernando Merces},
  title        = {Miner Malware Targets IoT, Offered in the Underground},
  year         = {2018},
  month        = may,
  day          = {2},
  howpublished = {\url{https://www.trendmicro.com/en_us/research/18/e/cryptocurrency-mining-malware-targeting-iot-being-offered-in-the-underground.html}},
  note         = {Trend Micro Research, accessed 2026-01-26}
}

@misc{kaspersky2022_crypto_miners_on_the_rise,
  author       = {Kaspersky},
  title        = {Crypto Miners on the Rise: Kaspersky Experts Report More than 230\% Growth in the Number of Malicious Mining Programs},
  year         = {2022},
  month        = nov,
  day          = {10},
  howpublished = {\url{https://www.kaspersky.com/about/press-releases/crypto-miners-on-the-rise-kaspersky-experts-report-more-than-230-growth-in-the-number-of-malicious-mining-programs}},
  note         = {Kaspersky press release, accessed 2026-01-26}
}

@misc{HitsxxWhatsminerUpgrade,
  author       = {{Hitsxx}},
  title        = {{Whatsminer Firmware Upgrade Toolkit}},
  year         = {2018},
  howpublished = {\url{https://github.com/Hitsxx/Whatsminer}},
  note         = {Public GitHub repository providing firmware packaging and remote upgrade scripts for Whatsminer ASIC miners}
}

@techreport{enisa2019_threat_landscape_2018,
  author      = {{European Union Agency for Cybersecurity (ENISA)}},
  title       = {ENISA Threat Landscape Report 2018: 15 Top Cyberthreats and Trends},
  institution = {ENISA},
  year        = {2019},
  month       = jan,
  url         = {https://www.enisa.europa.eu/sites/default/files/publications/WP2018%20O.1.2.1%20-%20ENISA%20Threat%20Landscape%202018.pdf},
  note        = {Accessed 2026-01-26}
}

@misc{NIST800193,
  author       = {{National Institute of Standards and Technology}},
  title        = {Platform Firmware Resiliency Guidelines},
  year         = {2018},
  howpublished = {\url{https://csrc.nist.gov/publications/detail/sp/800-193/final}}
}

@misc{ETSI303645,
  author       = {{ETSI}},
  title        = {Cyber Security for Consumer Internet of Things: Baseline Requirements},
  year         = {2020},
  howpublished = {\url{https://www.etsi.org/deliver/etsi_en/303600_303699/303645/02.01.01_60/en_303645v020101p.pdf}}
}

@misc{ENISA2019SupplyChain,
  author       = {{ENISA}},
  title        = {Good Practices for Supply Chain Cybersecurity},
  year         = {2019},
  howpublished = {\url{https://www.enisa.europa.eu/publications/good-practices-for-supply-chain-cybersecurity}}
}

@misc{NTIASSBOM2021,
  author       = {{National Telecommunications and Information Administration}},
  title        = {The Minimum Elements for a Software Bill of Materials (SBOM)},
  year         = {2021},
  howpublished = {\url{https://www.ntia.gov/report/2021/minimum-elements-software-bill-materials-sbom}}
}

@misc{Kuppusamy2017TUF,
  author       = {Ramya Kuppusamy and others},
  title        = {The Update Framework (TUF)},
  year         = {2017},
  howpublished = {USENIX Security Workshop},
  note         = {Workshop reference}
}

@misc{Uptane2019,
  author       = {Ramya Kuppusamy and others},
  title        = {Uptane: Securing Software Updates for Automobiles},
  year         = {2019},
  howpublished = {\url{https://uptane.org}}
}

@misc{NIST80092,
  author       = {{National Institute of Standards and Technology}},
  title        = {Guide to Computer Security Log Management},
  year         = {2006},
  howpublished = {\url{https://csrc.nist.gov/publications/detail/sp/800-92/final}}
}

@misc{queenant2016,
  author       = {Darren Pauli},
  title        = {Aussie Researcher Claims Antminer Bitcoin Devices Can Be Hijacked},
  year         = {2016},
  howpublished = {\url{https://www.theregister.com/2016/07/12/aussie_writes_app_to_hijack_scores_of_pricey_antmine_bitcoin_miners/}},
  note         = {Accessed 2026}
}

@misc{antbleed2017,
  author       = {Richard Chirgwin},
  title        = {Antminer Has Remote Shutdown Flaw (Antbleed)},
  year         = {2017},
  howpublished = {\url{https://www.theregister.com/2017/04/27/prospect_of_trouble_in_bitcoin_world_major_miner_vulnerable/}},
  note         = {Accessed 2026}
}

@misc{cve2018bitmain,
  author       = {{National Institute of Standards and Technology (NIST)}},
  title        = {CVE-2018-11220: Bitmain Antminer Remote Code Execution},
  year         = {2018},
  howpublished = {\url{https://nvd.nist.gov/vuln/detail/CVE-2018-11220}},
  note         = {NVD}
}

@misc{cve2022avalon,
  author       = {{National Institute of Standards and Technology (NIST)}},
  title        = {CVE-2022-36604: Canaan Avalon Authentication Bypass},
  year         = {2022},
  howpublished = {\url{https://nvd.nist.gov/vuln/detail/CVE-2022-36604}},
  note         = {NVD}
}

@misc{cve2022goldshell1,
  author       = {{National Institute of Standards and Technology (NIST)}},
  title        = {CVE-2022-24659: Goldshell Path Traversal},
  year         = {2022},
  howpublished = {\url{https://nvd.nist.gov/vuln/detail/CVE-2022-24659}},
  note         = {NVD}
}

@misc{cve2022goldshell2,
  author       = {{National Institute of Standards and Technology (NIST)}},
  title        = {CVE-2022-24660: Goldshell Debug Interface Exposure},
  year         = {2022},
  howpublished = {\url{https://nvd.nist.gov/vuln/detail/CVE-2022-24660}},
  note         = {NVD}
}

@misc{cve2022goldshell3,
  author       = {{National Institute of Standards and Technology (NIST)}},
  title        = {CVE-2022-24657: Goldshell Hardcoded Credentials},
  year         = {2022},
  howpublished = {\url{https://nvd.nist.gov/vuln/detail/CVE-2022-24657}},
  note         = {NVD}
}

@misc{chambers2022,
  author       = {James A. Chambers},
  title        = {Cryptocurrency ASIC Miners Security and Hacking Audit},
  year         = {2022},
  howpublished = {\url{https://jamesachambers.com/cryptocurrency-asic-miners-security-and-hacking-audit/}},
  note         = {Security analysis blog}
}

@misc{dcentral2025,
  author       = {{D-Central Technologies}},
  title        = {Infected ASICs: A Growing Menace for Crypto Miners},
  year         = {2025},
  howpublished = {\url{https://d-central.tech/infected-asics-a-growing-menace-for-crypto-miners-everywhere/}},
  note         = {Industry report}
}

@misc{bgp_hijack_mining,
  author       = {{Security Affairs}},
  title        = {Attacks on ISP Networks Allows to Steal \$83,000 from Bitcoin Mining Pools},
  year         = {2014},
  howpublished = {\url{https://securityaffairs.com/27448/cyber-crime/bitcoin-mining-pools-hack.html}},
  note         = {BGP hijacking of mining traffic}
}

@misc{satori_coin_robber,
  author       = {{Netlab 360}},
  title        = {Satori Coin Robber Malware Analysis},
  year         = {2018},
  howpublished = {\url{https://blog.netlab.360.com/botnets-never-die-satori-refuses-to-fade-away-en/}},
  note         = {Wallet replacement attack on mining software}
}

@misc{mirai_cloudflare,
  author       = {{Cloudflare}},
  title        = {Inside the Mirai Botnet},
  year         = {2016},
  howpublished = {\url{https://blog.cloudflare.com/inside-mirai-the-infamous-iot-botnet-a-retrospective-analysis/}},
  note         = {Large-scale IoT botnet and DDoS attacks}
}

@misc{hant_malware,
  author       = {{D-Central}},
  title        = {hAnt ASIC Malware Targeting Miners},
  year         = {2019},
  howpublished = {\url{https://d-central.tech/infected-asics-a-growing-menace-for-crypto-miners-everywhere/}},
  note         = {Ransomware targeting Antminer devices}
}

@misc{dd4bc_coindesk,
  author       = {{CoinDesk}},
  title        = {Bitcoin Mining Pools Targeted in Wave of DDoS Attacks},
  year         = {2015},
  howpublished = {\url{https://www.coindesk.com/markets/2015/03/12/bitcoin-mining-pools-targeted-in-wave-of-ddos-attacks}},
  note         = {DD4BC attacks on mining pools}
}

@misc{condibot_scmedia,
  author       = {{SC Media}},
  title        = {New Malware Targets Linux Devices for DDoS and Crypto Mining},
  year         = {2026},
  howpublished = {\url{https://www.scworld.com/brief/new-malware-targets-linux-network-devices-for-ddos-crypto-mining}},
  note         = {CondiBot and Monaco malware}
}

\clearpage
\appendices

\section{Open Science}
\label{app:openscience}

This work contributes a reproducible methodology for large-scale security analysis of cryptocurrency mining firmware based on publicly accessible distribution artifacts. To support open and verifiable research while avoiding legal and ethical risks, we do not redistribute proprietary firmware images.

We considered the following artifacts: vendor firmware update and flashing packages, unpacked firmware filesystems, collection and extraction scripts, static analysis tools, and validation configurations. Raw firmware binaries, unpacked filesystems, and end-to-end extraction pipelines are not publicly released. These artifacts are derived from production firmware distributed by major manufacturers and exhibit extensive code reuse across models and generations. Publishing them would materially lower the cost of exploitation at scale before mitigations are deployed.

Instead, we provide full methodological transparency. The paper details how to identify firmware distribution channels, collect publicly available artifacts, normalize formats, extract firmware contents, and perform static analysis without interacting with deployed systems. Reproducibility is achieved by reapplying the methodology rather than by reusing an identical dataset. The approach is vendor agnostic and transferable to other embedded and industrial firmware ecosystems.

For transparency, the paper cites the publicly available open-source repositories and tools that informed and supported our analysis. These references are provided in the bibliography.

\section{Extended comparison with related work}
\label{app:related_extended}

Prior work on firmware security and cryptocurrency mining follows two largely independent directions, which we detail below to highlight the gap addressed by our contribution.

Large-scale firmware analysis has demonstrated that publicly available firmware images enable scalable vulnerability discovery without requiring physical access to devices~\cite{CostinUsenix14}. This approach has been extended with dynamic analysis and emulation frameworks such as Firmadyne, Firm-AFL, and P2IM~\cite{costin2016dynamic,BaFIRMAFLUsenix19,FengP2IMUsenix20}, allowing deeper exploration of runtime behaviors and vulnerability classes. More recent surveys further systematize extraction techniques and common weaknesses in embedded firmware ecosystems~\cite{ulhaq2023firmware,bakhshi2024iotfirmwarevulnerabilities}. However, these works remain domain-agnostic and do not consider the specific constraints, attack surfaces, or economic incentives of cryptocurrency mining infrastructures. In particular, they do not model firmware update workflows as a primary attack vector, nor do they relate findings to concrete attacker objectives.

In parallel, prior work on cryptocurrency mining security has primarily focused on protocol and network-level threats. Several studies analyze weaknesses of the Stratum protocol, including man-in-the-middle attacks, hashrate redirection, and share manipulation~\cite{recabarren2017hardeningstratum,liu2021disappeared,StratumMITM,StratumAttackSurface}. Other works investigate routing-level attacks on mining pools or large-scale cryptojacking campaigns~\cite{Tran2024RoutingAttacksMiningPools,sari2017exploitingcryptocurrencyminers,zimba2021demystifyingcryptocurrencyminingattacks}. While these works provide valuable insights into network-level threats, they do not consider the firmware layer that governs update mechanisms, management interfaces, and device behavior, leaving a significant portion of the attack surface unexplored.

Public vulnerability disclosures further illustrate the existence of security weaknesses in real-world mining devices. Prior reports document backdoors and remote access mechanisms in Bitmain miners~\cite{antbleed2017,cve2018bitmain}, as well as authentication and API flaws affecting Avalon and Goldshell devices~\cite{cve2022avalon,cve2022goldshell1,cve2022goldshell2,cve2022goldshell3}. These findings consistently highlight recurring issues such as weak authentication, insecure update mechanisms, and exposed management interfaces. However, such disclosures remain fragmented, vendor-specific, and largely disconnected from a broader analysis of the attack surface. In particular, they do not capture the role of firmware distribution mechanisms as a primary entry point for large-scale attacks.

Overall, existing literature lacks a unified approach combining firmware-level analysis with mining-specific threat modeling. Firmware studies do not capture domain-specific attack implications, while mining security works overlook firmware distribution and update mechanisms. Existing vulnerability disclosures, although valuable, provide only partial visibility and do not explain how individual weaknesses compose into complete attack paths.

Our work addresses this gap by combining cross-vendor firmware analysis with mining-specific attack modeling. We rely on publicly distributed update artifacts as a reproducible evidence source, reconstruct internal architectures, and systematically map firmware-level weaknesses to concrete attacker capabilities and economic impact. This approach enables us to identify systemic design failures in the firmware lifecycle rather than isolated vulnerabilities.

A structured comparison with prior work is provided in Table~\ref{tab:related_comparison}, highlighting that our approach uniquely combines firmware analysis, mining-specific modeling, attack scenario construction, and cross-vendor coverage.

\begin{table}[t]
\centering
\tiny
\setlength{\tabcolsep}{3pt}
\begin{tabular}{|p{3.2cm}|c|c|c|c|}
\hline
\textbf{Work} & \textbf{Firmware} & \textbf{Mining-specific} & \textbf{Attack scenarios} & \textbf{Cross-vendor} \\
\hline
Costin et al.~\cite{CostinUsenix14} & Yes & No & No & Yes \\
\hline
Firmadyne~\cite{costin2016dynamic} & Yes & No & No & No \\
\hline
Firm-AFL~\cite{BaFIRMAFLUsenix19} & Yes & No & No & No \\
\hline
Stratum attacks~\cite{recabarren2017hardeningstratum,liu2021disappeared} & No & Yes & Yes & No \\
\hline
Routing attacks~\cite{Tran2024RoutingAttacksMiningPools} & No & Yes & Yes & No \\
\hline
\textbf{This work} & Yes & Yes & Yes & Yes \\
\hline
\end{tabular}
\caption{Comparison with related work}
\label{tab:related_comparison}
\end{table}

\section{Firmware architecture in details}\label{sec:minerarch}

We describe the firmware architecture of a cryptocurrency miner by decomposing it into its main software and hardware components, following the abstraction introduced in \Cref{fig:firmware_arch}.

\begin{itemize}

\item \textbf{Boot:} Early boot stage responsible for initializing the platform and transferring control to the system initialization logic.

\item \textbf{Init:} System initialization phase configuring hardware and preparing the runtime environment.

\item \textbf{Operating System (OS):} Embedded system (typically Linux) providing core services such as process management, networking, storage, and hardware abstraction.

\item \textbf{Libraries:} Shared components providing reusable functionalities, including cryptography, networking utilities, and system helpers.

\item \textbf{Miner:} Core application implementing the proof-of-work algorithm and orchestrating hashing operations on the hardware.

\item \textbf{Blockchain / Mining Pool:} External endpoint used to receive mining jobs and submit computed shares (e.g., via Stratum).

\item \textbf{Web Server:} Embedded interface exposing graphical monitoring and configuration capabilities.

\item \textbf{Control Console:} Optional interface enabling remote or fleet-level management.

\item \textbf{API:} Programmatic interface allowing automated configuration, monitoring, and integration with external systems.

\item \textbf{Firmware Update:} Mechanism responsible for installing firmware images or partial updates and enforcing the associated trust model.

\item \textbf{Debug Interfaces:} High-privilege access mechanisms, either software-based (e.g., SSH) or physical (e.g., UART), used for maintenance and diagnostics.

\item \textbf{Hardware Components:} Underlying platform including SoC, memory, networking interfaces, and sensors hosting the firmware stack.

\item \textbf{ASIC Chips:} Dedicated hardware executing hashing operations under the control of the miner software.

\end{itemize}

\section{Chronological Listing of ASIC Miners by Manufacturer}\label{sec:fulllist}
To improve readability and highlight temporal trends in the ASIC mining ecosystem, we group commercially available mining devices by manufacturer and release year, listing model names without algorithmic or performance details. Models are color-coded according to the most impactful attack scenario achievable under a \gls{lan} threat model, following a worst-case ordering
(\textcolor{pastelRed}{red} $>$ \textcolor{pastelOrange}{orange} $>$ \textcolor{pastelYellow}{yellow} $>$ \textcolor{pastelBlue}{blue} $>$ black) :
\textcolor{pastelRed}{full device takeover},
\textcolor{pastelOrange}{cryptocurrency theft and revenue redirection},
\textcolor{pastelYellow}{physical degradation and hardware damage}, and
\textcolor{pastelBlue}{performance degradation and operational disruption}.
This classification is restricted to attackers with network access to the miner environment. Outside this model, all devices are exposed to firmware phishing and plaintext Stratum~V1 attacks, which enable at least \textcolor{pastelOrange}{revenue redirection}, and would therefore place all miners in the orange category at least. As a reminder, this categorization by scenario should be taken with caution since it is not impossible that the vulnerabilities found in the update packages are not present or are disabled on the real models. \\

{\fontsize{8.3pt}{10pt}\selectfont
\noindent\textbf{Bitmain Antminer.}

\textbf{2026:}
U3S23H.

\textbf{2025:}
U2L9H,
\textcolor{pastelRed}{S21 Immersion 239T},
L11 Hyd 2U,
\textcolor{pastelRed}{L9 Hyd 2U},
L11 Hyd 6U,
L11 Pro,
L11,
\textcolor{pastelRed}{S21e Hyd (310Th)},
S23 Hyd 3U,
S23 Hyd,
S23 Immersion,
S23,
KS7 40T,
\textcolor{pastelRed}{S21e Hyd},
\textcolor{pastelRed}{S19 XP+ Hyd},
\textcolor{pastelRed}{S21+ Hydro},
\textcolor{pastelRed}{S21 XP+ Hyd},
\textcolor{pastelRed}{E11}.

\textbf{2024:}
\textcolor{pastelRed}{S21E XP HYD 430T},
\textcolor{pastelRed}{S21 XP Hydro},
\textcolor{pastelRed}{S21+},
\textcolor{pastelRed}{S21 Immersion 300T},
\textcolor{pastelRed}{S21E XP Hyd 3U},
\textcolor{pastelRed}{S21 Pro},
\textcolor{pastelRed}{AL1 Pro},
\textcolor{pastelRed}{AL1},
\textcolor{pastelRed}{S21 XP},
\textcolor{pastelRed}{DR7},
\textcolor{pastelRed}{L9 17.6Gh},
\textcolor{pastelRed}{L9 16Gh},
\textcolor{pastelRed}{KS5 Pro},
\textcolor{pastelRed}{KS5},
\textcolor{pastelRed}{S21 Hydro},
\textcolor{pastelRed}{S21}.

\textbf{2023:}
\textcolor{pastelRed}{T21},
\textcolor{pastelRed}{KS3},
\textcolor{pastelRed}{X5},
\textcolor{pastelRed}{S19K Pro (115Th)},
\textcolor{pastelRed}{S19j XP},
\textcolor{pastelRed}{Z15 Pro},
\textcolor{pastelRed}{S19k Pro},
\textcolor{pastelRed}{E9 Pro},
\textcolor{pastelRed}{D9}.

\textbf{2022:}
\textcolor{pastelRed}{HS3},
\textcolor{pastelRed}{K7},
\textcolor{pastelRed}{S19j Pro+},
\textcolor{pastelRed}{S19 XP Hydro},
\textcolor{pastelRed}{KA3},
\textcolor{pastelRed}{E9},
\textcolor{pastelRed}{S19 XP},
\textcolor{pastelRed}{S19 Pro+ Hyd},
\textcolor{pastelRed}{S19 Pro Hyd},
\textcolor{pastelRed}{L7}.

\textbf{2021:}
\textcolor{pastelRed}{Z15e},
\textcolor{pastelRed}{L7 9500Mh},
\textcolor{pastelRed}{L7 9300Mh},
\textcolor{pastelRed}{L7 9160Mh},
\textcolor{pastelRed}{L7 9050Mh},
\textcolor{pastelRed}{DR5 (35Th)},
\textcolor{pastelRed}{D7},
\textcolor{pastelRed}{S19j Pro (96Th)},
\textcolor{pastelRed}{S19j Pro (104Th)},
\textcolor{pastelRed}{S19j Pro (110Th)},
\textcolor{pastelRed}{S19j Pro (100Th)},
\textcolor{pastelRed}{S19j (90Th)}.

\textbf{2020:}
\textcolor{pastelRed}{Z15},
\textcolor{pastelRed}{T19},
\textcolor{pastelRed}{S19 Pro},
\textcolor{pastelRed}{S19 (95Th)},
\textcolor{pastelRed}{K5},
\textcolor{pastelRed}{T17+ 55Th},
\textcolor{pastelRed}{S17 Pro 59Th},
\textcolor{pastelRed}{S17+ 67Th},
\textcolor{pastelRed}{T17+ 58Th},
\textcolor{pastelRed}{T17 42Th},
\textcolor{pastelRed}{S17+ 70Th}.

\textbf{2019:}
\textcolor{pastelRed}{T17+ 64Th},
\textcolor{pastelRed}{S17+ 73Th},
\textcolor{pastelRed}{S17+ 76Th},
\textcolor{pastelRed}{S9 SE 17Th},
\textcolor{pastelRed}{T17e},
\textcolor{pastelRed}{S17e 64T},
\textcolor{pastelRed}{S17e 60T},
\textcolor{pastelRed}{S17 59Th},
\textcolor{pastelRed}{Z11J},
\textcolor{pastelRed}{Z11E},
\textcolor{pastelRed}{S9k},
\textcolor{pastelRed}{S9 SE 16Th},
\textcolor{pastelRed}{S17 Pro 56Th},
\textcolor{pastelRed}{T17 40Th},
\textcolor{pastelRed}{S17 Pro 50Th},
\textcolor{pastelRed}{S17 Pro 53Th},
\textcolor{pastelRed}{S17 56Th},
\textcolor{pastelRed}{S17 53Th},
\textcolor{pastelRed}{Z11},
\textcolor{pastelRed}{B7}.

\textbf{2018:}
\textcolor{pastelRed}{DR5 (34Th)},
\textcolor{pastelRed}{T15 (eco mode)},
\textcolor{pastelRed}{T15},
\textcolor{pastelRed}{S15 (eco mode)},
\textcolor{pastelRed}{S15},
\textcolor{pastelRed}{D5},
\textcolor{pastelRed}{S11},
\textcolor{pastelRed}{Z9},
\textcolor{pastelRed}{S9j 14.5T},
\textcolor{pastelRed}{S9 Hydro},
\textcolor{pastelRed}{E3},
\textcolor{pastelRed}{S9i},
\textcolor{pastelRed}{L3++},
\textcolor{pastelRed}{B3},
\textcolor{pastelRed}{X3},
\textcolor{pastelRed}{V9},
\textcolor{pastelRed}{A3},
\textcolor{pastelRed}{T9+}.

\textbf{2017 and earlier:}
\textcolor{pastelRed}{L3+},
\textcolor{pastelRed}{S9},
\textcolor{pastelRed}{T9},
\textcolor{pastelRed}{R4},
\textcolor{pastelRed}{S7-LN},
\textcolor{pastelRed}{S7},
\textcolor{pastelRed}{S5},
\textcolor{pastelRed}{S3}.

\smallskip
\smallskip
\noindent\textbf{MicroBT WhatsMiner.}

\textbf{2025:}
\textcolor{pastelBlue}{M79S},
\textcolor{pastelBlue}{M78S},
\textcolor{pastelBlue}{M76S+},
\textcolor{pastelBlue}{M76S},
\textcolor{pastelBlue}{M76},
\textcolor{pastelBlue}{M73S+},
\textcolor{pastelBlue}{M73S},
\textcolor{pastelBlue}{M73},
\textcolor{pastelBlue}{M72S},
\textcolor{pastelBlue}{M70S+},
\textcolor{pastelBlue}{M70S},
\textcolor{pastelBlue}{M70}.

\textbf{2024:}
\textcolor{pastelBlue}{M66S++},
\textcolor{pastelBlue}{M63S++},
\textcolor{pastelBlue}{M60S++},
\textcolor{pastelBlue}{M61S},
\textcolor{pastelBlue}{M66S+},
\textcolor{pastelBlue}{M63S+},
\textcolor{pastelBlue}{M60S+}.

\textbf{2023:}
\textcolor{pastelBlue}{M66S},
\textcolor{pastelBlue}{M66S Hydro},
\textcolor{pastelBlue}{M66},
\textcolor{pastelBlue}{M66 Hydro},
\textcolor{pastelBlue}{M66s},
\textcolor{pastelBlue}{M63S},
\textcolor{pastelBlue}{M63S Hydro},
\textcolor{pastelBlue}{M63},
\textcolor{pastelBlue}{M63 Hydro},
\textcolor{pastelBlue}{M63s},
\textcolor{pastelBlue}{M60},
\textcolor{pastelBlue}{M60S},
\textcolor{pastelBlue}{M53S+},
\textcolor{pastelBlue}{M56},
\textcolor{pastelBlue}{M56S},
\textcolor{pastelBlue}{M56 Hydro},
\textcolor{pastelBlue}{M36S+},
\textcolor{pastelBlue}{M53s},
\textcolor{pastelBlue}{M53},
\textcolor{pastelBlue}{M53S Hydro},
\textcolor{pastelBlue}{M53 Hydro},
\textcolor{pastelBlue}{M50s++}.

\textbf{2022:}
\textcolor{pastelBlue}{M33s++},
\textcolor{pastelBlue}{M50},
\textcolor{pastelBlue}{M50s}.

\textbf{2021 and earlier:}
\textcolor{pastelBlue}{M32s},
\textcolor{pastelBlue}{M32},
\textcolor{pastelBlue}{M31s+},
\textcolor{pastelBlue}{M31s},
\textcolor{pastelBlue}{M30s+},
\textcolor{pastelBlue}{M30s++},
\textcolor{pastelBlue}{M30s},
\textcolor{pastelBlue}{M20 45T},
\textcolor{pastelBlue}{M20S 65T},
\textcolor{pastelBlue}{M20S 68T},
\textcolor{pastelBlue}{M20s},
\textcolor{pastelBlue}{M21},
\textcolor{pastelBlue}{M21S 56T},
\textcolor{pastelBlue}{M21S 58T},
\textcolor{pastelBlue}{M21s},
\textcolor{pastelBlue}{D1},
\textcolor{pastelBlue}{M10},
\textcolor{pastelBlue}{M10s},
\textcolor{pastelBlue}{M10S},
\textcolor{pastelBlue}{M3X},
\textcolor{pastelBlue}{M3}.

\smallskip
\noindent\textbf{Canaan Avalon.}

\textbf{2026:}
A16XP-300T,
A16-282T.

\textbf{2025:}
\textcolor{pastelRed}{A1566HA 2U},
\textcolor{pastelRed}{Q},
\textcolor{pastelRed}{A15Pro-221T},
\textcolor{pastelRed}{A15 Pro 218T},
\textcolor{pastelRed}{Nano 3S},
\textcolor{pastelRed}{Mini 3}.

\textbf{2024:}
\textcolor{pastelRed}{A15XP 206T},
\textcolor{pastelRed}{1566},
\textcolor{pastelRed}{A1566I},
\textcolor{pastelOrange}{A1366I},
\textcolor{pastelRed}{Nano 3}.

\textbf{2023:}
\textcolor{pastelOrange}{1466},
\textcolor{pastelOrange}{A1446}.

\textbf{2022:}
\textcolor{pastelOrange}{1366},
\textcolor{pastelOrange}{1346}.

\textbf{2021:}
\textcolor{pastelOrange}{1126 Pro},
\textcolor{pastelOrange}{1246 85T},
\textcolor{pastelOrange}{1246 (90Th)}.

\textbf{2020:}
\textcolor{pastelRed}{1246 96T},
\textcolor{pastelRed}{1166 PRO 78Th},
\textcolor{pastelRed}{1166 PRO 72Th},
\textcolor{pastelRed}{1146 Pro},
\textcolor{pastelRed}{1146},
\textcolor{pastelRed}{1166}.

\textbf{2019:}
\textcolor{pastelRed}{1066},
\textcolor{pastelRed}{1047},
\textcolor{pastelRed}{911},
\textcolor{pastelRed}{852}.

\textbf{2018:}
\textcolor{pastelRed}{921},
\textcolor{pastelRed}{1126},
\textcolor{pastelRed}{841},
\textcolor{pastelRed}{821}.

\textbf{2017 and earlier:}
\textcolor{pastelRed}{741},
\textcolor{pastelRed}{1}.

\smallskip
\noindent\textbf{IceRiver.}

\textbf{2025:}
\textcolor{pastelOrange}{ALEO AE3},
\textcolor{pastelOrange}{ALEO AE2},
\textcolor{pastelOrange}{ALEO AE1 Lite (250MH)},
\textcolor{pastelOrange}{ALEO AE1 Lite (300MH)},
\textcolor{pastelOrange}{ALEO AE0},
\textcolor{pastelOrange}{KAS KS7},
\textcolor{pastelOrange}{KAS KS7 Lite}.

\textbf{2024:}
\textcolor{pastelOrange}{KS2 Lite},
\textcolor{pastelOrange}{AL2 Lite},
\textcolor{pastelOrange}{AL3},
\textcolor{pastelOrange}{RX0},
\textcolor{pastelOrange}{ALPH AL0},
\textcolor{pastelOrange}{KS0 Ultra},
\textcolor{pastelOrange}{KS5M},
\textcolor{pastelOrange}{KS5L}.

\textbf{2023:}
\textcolor{pastelOrange}{KS0 Pro},
\textcolor{pastelOrange}{KS1},
\textcolor{pastelOrange}{KS2},
\textcolor{pastelOrange}{KS3},
\textcolor{pastelOrange}{KS3L},
\textcolor{pastelOrange}{KS3M},
\textcolor{pastelOrange}{KS0}.

}

\section{Deeper Analysis of Recent Firmware}
\label{sec:deeper_analysis}

\begin{table}[ht!]
\centering
\footnotesize
\setlength{\tabcolsep}{3pt}
\renewcommand{\arraystretch}{1.1}
\begin{tabularx}{\columnwidth}{p{0.56\columnwidth} >{\raggedright\ttfamily\arraybackslash}X}
\toprule
\textbf{Issue} & \textbf{File} \\
\midrule
Partial signature verification of update artifacts. &
sbin/updateproc.sh \\

Unsalted MD5 password hashing. &
www/pages/cgi-bin/passwd.cgi \newline
lib/lighttpd/mod\_auth.so \\

Runtime-enabled Dropbear SSH service. &
etc/init.d/S50dropbear \\

Legacy services (Telnet, FTP, TFTP) supported. &
etc/services \\

Status endpoint leaks runtime information. &
lib/lighttpd/mod\_status.so \\

Cookies missing \texttt{HttpOnly}. &
lib/lighttpd/mod\_usertrack.so \\

Legacy SSH algorithms enabled. &
sbin/dropbear \\
\bottomrule
\end{tabularx}
\caption{Vulnerabilities identified by static analysis in Bitmain miners.}
\label{tab:vuln_per_file_bitmain_recent}
\end{table}

\begin{table}[ht!]
\centering
\footnotesize
\setlength{\tabcolsep}{3pt}
\renewcommand{\arraystretch}{1.1}
\begin{tabularx}{\columnwidth}{p{0.56\columnwidth} >{\raggedright\ttfamily\arraybackslash}X}
\toprule
\textbf{Issue} & \textbf{File} \\
\midrule
SSH service enabled at boot time. &
usr/sbin/sshd \newline
etc/init.d/S50sshd \\

Default SSH configuration with limited hardening. &
etc/ssh/sshd\_config \\

Authentication secrets stored in the root filesystem. &
etc/shadow \\

OpenSSH linked against legacy OpenSSL 1.1.1. &
usr/sbin/sshd \newline
usr/lib/libssl.so.1.1 \\

Privileged CGI actions without explicit CSRF protection. &
administrator.html \newline
reboot.html \newline
networkcfg.html \\
\bottomrule
\end{tabularx}
\caption{Vulnerabilities identified by static analysis in Canaan miners.}
\label{tab:vuln_per_file_canaan_recent}
\end{table}

This section presents a deeper, targeted analysis of cryptocurrency miner firmware, focusing exclusively on recent and actively deployed models. While the previous sections provide a large and longitudinal view of the ecosystem, the goal here is to evaluate the current security posture of one modern devices per manufacturer.

\subsection{Bitmain}
\label{subsec:bitmain}

We begin with Bitmain by analyzing a recent firmware image,
{\fontsize{10.15pt}{12pt}\selectfont\texttt{Bitmain\_2025-11-19\_FR-1.27\_251009-S19XP\_2B\_Hyd}}\hfill, 
corresponding to an Antminer S19 XP Hyd device.

We first perform a file-level static analysis to identify concrete implementation weaknesses affecting update verification, authentication mechanisms, and exposed services.
The resulting vulnerabilities are summarized in \Cref{tab:vuln_per_file_bitmain_recent}.

In addition, we extract and enumerate the main embedded software components and their versions in order to characterize the underlying software stack and assess its reliance on legacy components:

\vspace{3mm}
\newcolumntype{L}[1]{>{\raggedright\arraybackslash}p{#1}}
\begin{tabular}{@{}L{0.43\linewidth} L{0.48\linewidth}@{}}
\textbullet\ \textbf{lighttpd}~1.4.32       & \textbullet\ \textbf{libstdc++}~6.0.24      \\[1mm]
\textbullet\ \textbf{BusyBox}~v1.29.3       & \textbullet\ \textbf{U-Boot}~2016.07        \\[1mm]
\textbullet\ \textbf{Dropbear SSH}~2018.76  & \textbullet\ \textbf{OpenSSL}~1.1.1 (early) \\[1mm]
\textbullet\ \textbf{GNU libc (glibc)}~2.25 & \\
\end{tabular}

\subsection{Canaan}
\label{subsec:canaan}

We analyze a Canaan Avalon 15 firmware update package : \\
Canaan\_Avalon\_15xHY\_release\_OTA\_2025111202\_773bb92.aup \\
The vulnerabilities are summarized in 
\Cref{tab:vuln_per_file_canaan_recent}.
The embedded software stack includes:

\vspace{3mm}
\begin{tabular}{@{}L{0.43\linewidth} L{0.48\linewidth}@{}}
\textbullet\ \textbf{BusyBox}~v1.33.0       & \textbullet\ \textbf{OpenSSH (sshd)}~8.3p1     \\[1mm]
\textbullet\ \textbf{GNU libc (glibc)}~2.33 & \textbullet\ \textbf{OpenSSL}~1.1.1i           \\[1mm]
\multicolumn{2}{@{}l@{}}{\textbullet\ \textbf{libstdc++}~GLIBCXX\_3.4.28 (CXXABI\_1.3.12)}         \\
\end{tabular}

\begin{table}[t]
\centering
\footnotesize
\setlength{\tabcolsep}{3pt}
\renewcommand{\arraystretch}{1.1}
\begin{tabularx}{\columnwidth}{p{0.56\columnwidth} >{\raggedright\ttfamily\arraybackslash}X}
\toprule
\textbf{Issue} & \textbf{File} \\
\midrule

SSH activation via temporary file trigger. &
/tmp/dropbear\_on \\

SSH config allows password auth and root login. &
etc/config/dropbear \\

MD5 password hash with weak defaults. &
etc/shadow \\

Bootloader integrity verified via external checksum only. &
etc/uboot.md5 \\

\bottomrule
\end{tabularx}
\caption{Vulnerabilities identified by static analysis in MicroBT miners.}
\label{tab:vuln_per_file_microbt_recent}
\end{table}

\begin{table}[t]
\centering
\footnotesize
\setlength{\tabcolsep}{3pt}
\renewcommand{\arraystretch}{1.1}
\begin{tabularx}{\columnwidth}{p{0.56\columnwidth} >{\raggedright\ttfamily\arraybackslash}X}
\toprule
\textbf{Issue} & \textbf{File} \\
\midrule

Legacy password hashing (unsalted MD5). &
bg/linux-*/bin/authpass \\

HTTP Digest authentication using MD5. &
bg/linux-*/obj/httpLib.o \\

Cookies missing \texttt{HttpOnly}. &
bg/web/js/login.js \\

Client-side dynamic code execution via \texttt{eval}. &
bg/web/js/translate.js \\

Unsafe C string functions (\texttt{sprintf}, \texttt{strcpy}). &
bg/controllers/user.c \\

\bottomrule
\end{tabularx}
\caption{Vulnerabilities identified by static analysis in Iceriver miners.}
\label{tab:vuln_per_file_iceriver_recent}
\end{table}

\subsection{MicroBT}
\label{subsec:microbt}

We analyze a MicroBT firmware flash image corresponding to a WhatsMiner device.
This analysis is restricted to flash-resident firmware components.
Parts of the firmware image are encrypted, and additional software components are provisioned later through the update mechanism.

The vulnerabilities are summarized in
\Cref{tab:vuln_per_file_microbt_recent}.

The flash-resident firmware contains configuration files and initialization scripts for several system components.
However, executable binaries for core libraries and services (e.g., SSH daemon, cryptographic libraries, standard C library) are not present in the extracted flash image.
As a result, component versions cannot be reliably recovered through static analysis.
We deliberately avoid inferring versions from indirect traces or configuration artifacts.

\subsection{Iceriver}
\label{subsec:iceriver}

We analyze Iceriver through a static analysis of the embedded web management stack, the only firmware component available as a standalone artifact.

The vulnerabilities are summarized in \Cref{tab:vuln_per_file_iceriver_recent}.

To characterize software obsolescence, we enumerate the runtime components and versions explicitly present in the web stack:

\begin{itemize}
  \item \textbf{Embedthis Appweb}~4.x (embedded web server and ESP framework)
  \item \textbf{glibc}~2.2.5--2.4 (legacy symbol versions referenced by binaries)
  \item \textbf{Jansson}~4.11.0 (\texttt{libjansson.so.4.11.0})
\end{itemize}

\begin{figure*}[t]
  \centering
  \includegraphics[width=\textwidth]{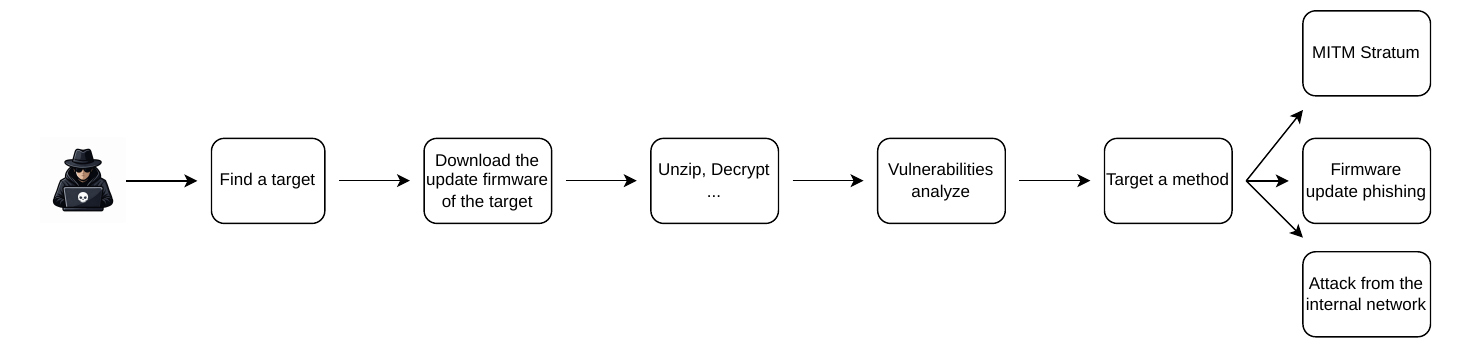}
  \caption{Attacker pipeline from initial access to final objectives.}
  \label{fig:pipelineattacker}
\end{figure*}

\begin{figure*}[t]
  \centering
  \includegraphics[width=\textwidth]{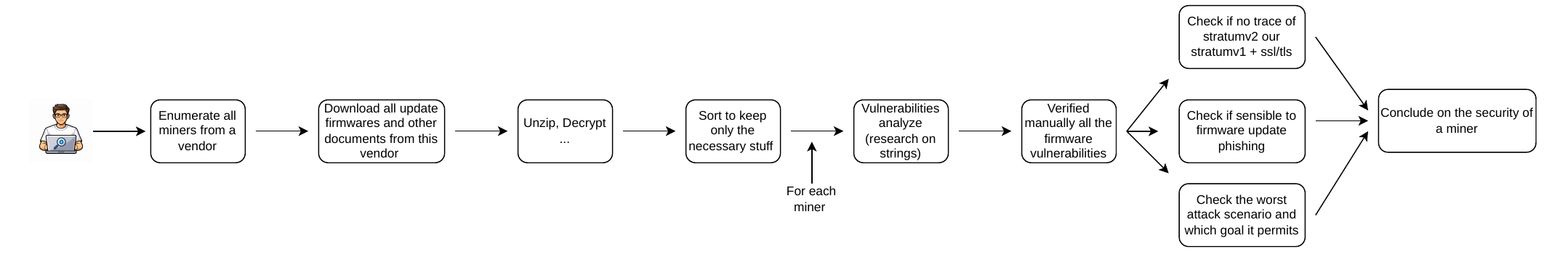}
  \caption{Analysis pipeline.}
  \label{fig:our_pipeline}
\end{figure*}

\end{document}